%
%
%
\def\unredoffs{} 

%
%
%
%
\newbox\leftpage \newdimen\fullhsize \newdimen\hstitle \newdimen\hsbody
\tolerance=1000\hfuzz=2pt
\catcode`\@=11 
\magnification=1200\unredoffs\baselineskip=16pt plus 2pt minus 1pt
\hsbody=\hsize \hstitle=\hsize 
%
%
%
%
%
\newcount\yearltd\yearltd=\year\advance\yearltd by -1900

\def\Title#1#2{\nopagenumbers\abstractfont\hsize=\hstitle\rightline{#1}%
\vskip 1in\centerline{\titlefont #2}\abstractfont\vskip .5in\pageno=0}
\def\Date#1{\vfill\leftline{#1}\tenpoint\supereject\global\hsize=\hsbody%
\footline={\hss\tenrm\folio\hss}}
%

\def\draftmode{\message{ DRAFTMODE }\def\draftdate{{\rm preliminary draft:
\number\month/\number\day/\number\yearltd\ \ \hourmin}}%
\headline={\hfil\draftdate}\writelabels\baselineskip=20pt plus 2pt minus 2pt
 {\count255=\time\divide\count255 by 60 \xdef\hourmin{\number\count255}
  \multiply\count255 by-60\advance\count255 by\time
  \xdef\hourmin{\hourmin:\ifnum\count255<10 0\fi\the\count255}}}
\def\nolabels{\def\wrlabeL##1{}\def\eqlabeL##1{}\def\reflabeL##1{}}
\def\writelabels{\def\wrlabeL##1{\leavevmode\vadjust{\rlap{\smash%
{\line{{\escapechar=` \hfill\rlap{\sevenrm\hskip.03in\string##1}}}}}}}%
\def\eqlabeL##1{{\escapechar-1\rlap{\sevenrm\hskip.05in\string##1}}}%
\def\reflabeL##1{\noexpand\llap{\noexpand\sevenrm\string\string\string##1}}}
\nolabels
%
\global\newcount\secno \global\secno=0
\global\newcount\meqno \global\meqno=1
\def\newsec#1{\global\advance\secno by1\message{(\the\secno. #1)}
\global\subsecno=0\eqnres@t\noindent{\bf\the\secno. #1}
\writetoca{{\secsym} {#1}}\par\nobreak\medskip\nobreak}
\def\eqnres@t{\xdef\secsym{\the\secno.}\global\meqno=1\bigbreak\bigskip}
\def\sequentialequations{\def\eqnres@t{\bigbreak}}\xdef\secsym{}
\global\newcount\subsecno \global\subsecno=0
\def\subsec#1{\global\advance\subsecno by1\message{(\secsym\the\subsecno. #1)}
\ifnum\lastpenalty>9000\else\bigbreak\fi
\noindent{\it\secsym\the\subsecno. #1}\writetoca{\string\quad 
{\secsym\the\subsecno.} {#1}}\par\nobreak\medskip\nobreak}
\def\appendix#1#2{\global\meqno=1\global\subsecno=0\xdef\secsym{\hbox{#1.}}
\bigbreak\bigskip\noindent{\bf Appendix #1. #2}\message{(#1. #2)}
\writetoca{Appendix {#1.} {#2}}\par\nobreak\medskip\nobreak}
%
%
\def\eqnn#1{\xdef #1{(\secsym\the\meqno)}\writedef{#1\leftbracket#1}%
\global\advance\meqno by1\wrlabeL#1}
\def\eqna#1{\xdef #1##1{\hbox{$(\secsym\the\meqno##1)$}}
\writedef{#1\numbersign1\leftbracket#1{\numbersign1}}%
\global\advance\meqno by1\wrlabeL{#1$\{\}$}}
\def\eqn#1#2{\xdef #1{(\secsym\the\meqno)}\writedef{#1\leftbracket#1}%
\global\advance\meqno by1$$#2\eqno#1\eqlabeL#1$$}
%
\newskip\footskip\footskip14pt plus 1pt minus 1pt 
\def\footnotefont{\ninepoint}\def\f@t#1{\footnotefont #1\@foot}
\def\f@@t{\baselineskip\footskip\bgroup\footnotefont\aftergroup\@foot\let\next}
\setbox\strutbox=\hbox{\vrule height9.5pt depth4.5pt width0pt}
\global\newcount\ftno \global\ftno=0
\def\foot{\global\advance\ftno by1\footnote{$^{\the\ftno}$}}
%
\newwrite\ftfile   
\def\footend{\def\foot{\global\advance\ftno by1\chardef\wfile=\ftfile
$^{\the\ftno}$\ifnum\ftno=1\immediate\openout\ftfile=foots.tmp\fi%
\immediate\write\ftfile{\noexpand\smallskip%
\noexpand\item{f\the\ftno:\ }\pctsign}\findarg}%
\def\footatend{\vfill\eject\immediate\closeout\ftfile{\parindent=20pt
\centerline{\bf Footnotes}\nobreak\bigskip\input foots.tmp }}}
\def\footatend{}
%
%
\global\newcount\refno \global\refno=1
\newwrite\rfile
\def\ref{[\the\refno]\nref}
\def\nref#1{\xdef#1{[\the\refno]}\writedef{#1\leftbracket#1}%
\ifnum\refno=1\immediate\openout\rfile=refs.tmp\fi
\global\advance\refno by1\chardef\wfile=\rfile\immediate
\write\rfile{\noexpand\item{#1\ }\reflabeL{#1\hskip.31in}\pctsign}\findarg}
\def\findarg#1#{\begingroup\obeylines\newlinechar=`\^^M\pass@rg}
{\obeylines\gdef\pass@rg#1{\writ@line\relax #1^^M\hbox{}^^M}%
\gdef\writ@line#1^^M{\expandafter\toks0\expandafter{\striprel@x #1}%
\edef\next{\the\toks0}\ifx\next\em@rk\let\next=\endgroup\else\ifx\next\empty%
\else\immediate\write\wfile{\the\toks0}\fi\let\next=\writ@line\fi\next\relax}}
\def\striprel@x#1{} \def\em@rk{\hbox{}} 
\def\lref{\begingroup\obeylines\lr@f}
\def\lr@f#1#2{\gdef#1{\ref#1{#2}}\endgroup\unskip}

\def\addref#1{\immediate\write\rfile{\noexpand\item{}#1}} 
\def\startrefs#1{\immediate\openout\rfile=refs.tmp\refno=#1}
\def\xref{\expandafter\xr@f}\def\xr@f[#1]{#1}
\def\refs#1{\count255=1[\r@fs #1{\hbox{}}]}
\def\r@fs#1{\ifx\und@fined#1\message{reflabel \string#1 is undefined.}%
\nref#1{need to supply reference \string#1.}\fi%
\vphantom{\hphantom{#1}}\edef\next{#1}\ifx\next\em@rk\def\next{}%
\else\ifx\next#1\ifodd\count255\relax\xref#1\count255=0\fi%
\else#1\count255=1\fi\let\next=\r@fs\fi\next}
%

%
\newwrite\ffile\global\newcount\figno \global\figno=1
\def\fig{fig.~\the\figno\nfig}
\def\nfig#1{\xdef#1{fig.~\the\figno}%
\writedef{#1\leftbracket fig.\noexpand~\the\figno}%
\ifnum\figno=1\immediate\openout\ffile=figs.tmp\fi\chardef\wfile=\ffile%
\immediate\write\ffile{\noexpand\medskip\noexpand\item{Fig.\ \the\figno. }
\reflabeL{#1\hskip.55in}\pctsign}\global\advance\figno by1\findarg}
\def\xfig{\expandafter\xf@g}\def\xf@g fig.\penalty\@M\ {}
\def\figs#1{figs.~\f@gs #1{\hbox{}}}
\def\f@gs#1{\edef\next{#1}\ifx\next\em@rk\def\next{}\else
\ifx\next#1\xfig #1\else#1\fi\let\next=\f@gs\fi\next}
\newwrite\lfile
{\escapechar-1\xdef\pctsign{\string\%}\xdef\leftbracket{\string\{}
\xdef\rightbracket{\string\}}\xdef\numbersign{\string\#}}

\def\writestop{\def\writestoppt{\immediate\write\lfile{\string\pageno%
\the\pageno\string\startrefs\leftbracket\the\refno\rightbracket%
\string\def\string\secsym\leftbracket\secsym\rightbracket%
\string\secno\the\secno\string\meqno\the\meqno}\immediate\closeout\lfile}}
\def\writestoppt{}\def\writedef#1{}
\def\seclab#1{\xdef #1{\the\secno}\writedef{#1\leftbracket#1}\wrlabeL{#1=#1}}
\def\subseclab#1{\xdef #1{\secsym\the\subsecno}%
\writedef{#1\leftbracket#1}\wrlabeL{#1=#1}}
\newwrite\tfile \def\writetoca#1{}
\def\leaderfill{\leaders\hbox to 1em{\hss.\hss}\hfill}
\def\writetoc{\immediate\openout\tfile=toc.tmp 
   \def\writetoca##1{{\edef\next{\write\tfile{\noindent ##1 
   \string\leaderfill {\noexpand\number\pageno} \par}}\next}}}

%
\catcode`\@=12 
%
\edef\tfontsize{\ifx\answ\bigans scaled\magstep3\else scaled\magstep4\fi}
\font\titlerm=cmr10 \tfontsize \font\titlerms=cmr7 \tfontsize
\font\titlermss=cmr5 \tfontsize \font\titlei=cmmi10 \tfontsize
\font\titleis=cmmi7 \tfontsize \font\titleiss=cmmi5 \tfontsize
\font\titlesy=cmsy10 \tfontsize \font\titlesys=cmsy7 \tfontsize
\font\titlesyss=cmsy5 \tfontsize \font\titleit=cmti10 \tfontsize
\skewchar\titlei='177 \skewchar\titleis='177 \skewchar\titleiss='177
\skewchar\titlesy='60 \skewchar\titlesys='60 \skewchar\titlesyss='60
\def\titlefont{\def\rm{\fam0\titlerm}
\textfont0=\titlerm \scriptfont0=\titlerms \scriptscriptfont0=\titlermss
\textfont1=\titlei \scriptfont1=\titleis \scriptscriptfont1=\titleiss
\textfont2=\titlesy \scriptfont2=\titlesys \scriptscriptfont2=\titlesyss
\textfont\itfam=\titleit \def\it{\fam\itfam\titleit}\rm}
 \ifx\answ\bigans\else scaled\magstep1\fi
\ifx\answ\bigans\def\abstractfont{\tenpoint}\else
\font\abssl=cmsl10 scaled \magstep1
\font\absrm=cmr10 scaled\magstep1 \font\absrms=cmr7 scaled\magstep1
\font\absrmss=cmr5 scaled\magstep1 \font\absi=cmmi10 scaled\magstep1
\font\absis=cmmi7 scaled\magstep1 \font\absiss=cmmi5 scaled\magstep1
\font\abssy=cmsy10 scaled\magstep1 \font\abssys=cmsy7 scaled\magstep1
\font\abssyss=cmsy5 scaled\magstep1 \font\absbf=cmbx10 scaled\magstep1
\skewchar\absi='177 \skewchar\absis='177 \skewchar\absiss='177
\skewchar\abssy='60 \skewchar\abssys='60 \skewchar\abssyss='60
\def\abstractfont{\def\rm{\fam0\absrm}
\textfont0=\absrm \scriptfont0=\absrms \scriptscriptfont0=\absrmss
\textfont1=\absi \scriptfont1=\absis \scriptscriptfont1=\absiss
\textfont2=\abssy \scriptfont2=\abssys \scriptscriptfont2=\abssyss
\textfont\itfam=\bigit \def\it{\fam\itfam\bigit}\def\footnotefont{\tenpoint}%
\textfont\slfam=\abssl \def\sl{\fam\slfam\abssl}%
\textfont\bffam=\absbf \def\bf{\fam\bffam\absbf}\rm}\fi
\def\tenpoint{\def\rm{\fam0\tenrm}
\textfont0=\tenrm \scriptfont0=\sevenrm \scriptscriptfont0=\fiverm
\textfont1=\teni  \scriptfont1=\seveni  \scriptscriptfont1=\fivei
\textfont2=\tensy \scriptfont2=\sevensy \scriptscriptfont2=\fivesy
\textfont\itfam=\tenit \def\it{\fam\itfam\tenit}\def\footnotefont{\ninepoint}%
\textfont\bffam=\tenbf \def\bf{\fam\bffam\tenbf}\def\sl{\fam\slfam\tensl}\rm}
\font\ninerm=cmr9 \font\sixrm=cmr6 \font\ninei=cmmi9 \font\sixi=cmmi6 
\font\ninesy=cmsy9 \font\sixsy=cmsy6 \font\ninebf=cmbx9 
\font\nineit=cmti9 \font\ninesl=cmsl9 \skewchar\ninei='177
\skewchar\sixi='177 \skewchar\ninesy='60 \skewchar\sixsy='60 
\def\ninepoint{\def\rm{\fam0\ninerm}
\textfont0=\ninerm \scriptfont0=\sixrm \scriptscriptfont0=\fiverm
\textfont1=\ninei \scriptfont1=\sixi \scriptscriptfont1=\fivei
\textfont2=\ninesy \scriptfont2=\sixsy \scriptscriptfont2=\fivesy
\textfont\itfam=\ninei \def\it{\fam\itfam\nineit}\def\sl{\fam\slfam\ninesl}%
\textfont\bffam=\ninebf \def\bf{\fam\bffam\ninebf}\rm} 
%
%
\def\noblackbox{\overfullrule=0pt}
\hyphenation{anom-aly anom-alies coun-ter-term coun-ter-terms}
\def\inv{^{\raise.15ex\hbox{${\scriptscriptstyle -}$}\kern-.05em 1}}

\def\Dsl{\,\raise.15ex\hbox{/}\mkern-13.5mu D} 
\def\dsl{\raise.15ex\hbox{/}\kern-.57em\partial}

\font\bigit=cmti10 scaled \magstep1
\def\lspace{\ifx\answ\bigans{}\else\qquad\fi}
\def\lbspace{\ifx\answ\bigans{}\else\hskip-.2in\fi} 
\def\boxeqn#1{\vcenter{\vbox{\hrule\hbox{\vrule\kern3pt\vbox{\kern3pt
        \hbox{${\displaystyle #1}$}\kern3pt}\kern3pt\vrule}\hrule}}}
\def\mbox#1#2{\vcenter{\hrule \hbox{\vrule height#2in
                \kern#1in \vrule} \hrule}}  
%

\def\darr#1{\raise1.5ex\hbox{$\leftrightarrow$}\mkern-16.5mu #1}

\def\roughly#1{\raise.3ex\hbox{$#1$\kern-.75em\lower1ex\hbox{$\sim$}}}

\openup -1pt
\input epsf
\expandafter\ifx\csname pre amssym.tex at\endcsname\relax \else\endinput\fi
\expandafter\chardef\csname pre amssym.tex at\endcsname=\the\catcode`\@
\catcode`\@=11
\ifx\undefined\newsymbol \else \begingroup\def\input#1 {\endgroup}\fi
\expandafter\ifx\csname amssym.def\endcsname\relax \else\endinput\fi
\expandafter\edef\csname amssym.def\endcsname{%
       \catcode`\noexpand\@=\the\catcode`\@\space}
\catcode`\@=11
\def\undefine#1{\let#1\undefined}
\def\newsymbol#1#2#3#4#5{\let\next@\relax
 \ifnum#2=\@ne\let\next@\msafam@\else
 \ifnum#2=\tw@\let\next@\msbfam@\fi\fi
 \mathchardef#1="#3\next@#4#5}
\def\mathhexbox@#1#2#3{\relax
 \ifmmode\mathpalette{}{\m@th\mathchar"#1#2#3}%
 \else\leavevmode\hbox{$\m@th\mathchar"#1#2#3$}\fi}
\def\hexnumber@#1{\ifcase#1 0\or 1\or 2\or 3\or 4\or 5\or 6\or 7\or 8\or
 9\or A\or B\or C\or D\or E\or F\fi}
\font\tenmsa=msam10
\font\sevenmsa=msam7
\font\fivemsa=msam5
\newfam\msafam
\textfont\msafam=\tenmsa
\scriptfont\msafam=\sevenmsa
\scriptscriptfont\msafam=\fivemsa
\edef\msafam@{\hexnumber@\msafam}
\mathchardef\dabar@"0\msafam@39
\def\maltese{{\mathhexbox@\msafam@7A}}
\font\tenmsb=msbm10
\font\sevenmsb=msbm7
\font\fivemsb=msbm5
\newfam\msbfam
\textfont\msbfam=\tenmsb
\scriptfont\msbfam=\sevenmsb
\scriptscriptfont\msbfam=\fivemsb
\edef\msbfam@{\hexnumber@\msbfam}
\def\Bbb#1{{\fam\msbfam\relax#1}}
\def\widehat#1{\setbox\z@\hbox{$\m@th#1$}%
 \ifdim\wd\z@>\tw@ em\mathaccent"0\msbfam@5B{#1}%
 \else\mathaccent"0362{#1}\fi}
\def\widetilde#1{\setbox\z@\hbox{$\m@th#1$}%
 \ifdim\wd\z@>\tw@ em\mathaccent"0\msbfam@5D{#1}%
 \else\mathaccent"0365{#1}\fi}
\font\teneufm=eufm10
\font\seveneufm=eufm7
\font\fiveeufm=eufm5
\newfam\eufmfam
\textfont\eufmfam=\teneufm
\scriptfont\eufmfam=\seveneufm
\scriptscriptfont\eufmfam=\fiveeufm
\def\frak#1{{\fam\eufmfam\relax#1}}

\csname amssym.def\endcsname
\relax
\newsymbol\smallsetminus 2272
\noblackbox
\newcount\figno
\figno=0
\def\mathrm#1{{\rm #1}}
\def\fig#1#2#3{
\par\begingroup\parindent=0pt\leftskip=1cm\rightskip=1cm\parindent=0pt
\baselineskip=11pt
\global\advance\figno by 1
\midinsert
\epsfxsize=#3
\centerline{\epsfbox{#2}}
\vskip 12pt
\centerline{{\bf Figure \the\figno} #1}\par
\endinsert\endgroup\par}
\font\tenmsb=msbm10       \font\sevenmsb=msbm7
\font\fivemsb=msbm5       \newfam\msbfam
\textfont\msbfam=\tenmsb  \scriptfont\msbfam=\sevenmsb
\scriptscriptfont\msbfam=\fivemsb
\def\Bbb#1{{\fam\msbfam\relax#1}}

\def\Rop{{\Bbb R}}
\def\Zop{{\Bbb Z}}

\def\Nop{{\Bbb N}}

\def\bbbc{{\mathchoice {\setbox0=\hbox{$\displaystyle\rm C$}\hbox{\hbox
to0pt{\kern0.4\wd0\vrule height0.9\ht0\hss}\box0}}
{\setbox0=\hbox{$\textstyle\rm C$}\hbox{\hbox
to0pt{\kern0.4\wd0\vrule height0.9\ht0\hss}\box0}}
{\setbox0=\hbox{$\scriptstyle\rm C$}\hbox{\hbox
to0pt{\kern0.4\wd0\vrule height0.9\ht0\hss}\box0}}
{\setbox0=\hbox{$\scriptscriptstyle\rm C$}\hbox{\hbox
to0pt{\kern0.4\wd0\vrule height0.9\ht0\hss}\box0}}}}
\def\figlabel#1{\xdef#1{\the\figno}}
\def\pano{\par\noindent}

\def\pmb#1{\setbox0=\hbox{#1}%
 \kern-.025em\copy0\kern-\wd0
 \kern.05em\copy0\kern-\wd0
 \kern-.025em\raise.0433em\box0 }



\def\hsmallsetminus{\hbox{\raise1.5pt\hbox{$\smallsetminus$}}}
\def\tilM{\hbox{${\scriptstyle \widetilde{\phantom M}}$}\hskip-9pt
   \raise1.3pt\hbox{${\scriptstyle M}$}\,}
\def\ie{{\it i.e.}}

\def\det{{\rm det}}
%


\def\figin{\epsfcheck\figin}\def\figins{\epsfcheck\figins}
\def\epsfcheck{\ifx\epsfbox\UnDeFiNeD
\message{(NO epsf.tex, FIGURES WILL BE IGNORED)}
\gdef\figin##1{\vskip2in}\gdef\figins##1{\hskip.5in}
\else\message{(FIGURES WILL BE INCLUDED)}%
\gdef\figin##1{##1}\gdef\figins##1{##1}\fi}
\def\DefWarn#1{}
\def\figinsert{\goodbreak\midinsert}
\def\ifig#1#2#3{\DefWarn#1\xdef#1{fig.~\the\figno}
\writedef{#1\leftbracket fig.\noexpand~\the\figno}%
\figinsert\figin{\centerline{#3}}\medskip\centerline{\vbox{\baselineskip12pt
\advance\hsize by -1truein\noindent\footnotefont{\bf Fig.~\the\figno:} #2}}
\bigskip\endinsert\global\advance\figno by1}
\figno=1


\def\II{{\rm II}}
\def\Z{{\hbox{Z}}}
\def\ie{{\it i.e.}}

\def\g{{\frak g}}


\lref\pwone{P. West, {\it Hidden superconformal symmetry in M-theory},
J. High Energy Phys. {\bf 0008} (2000) 007; {\tt hep-th/0005270}.}

\lref\pwtwo{P. West, {\it $E_{11}$ and M-theory},
Class. Quant. Grav. {\bf 18} (2001) 4443; {\tt hep-th/0104081}.} 

\lref\go{P. Goddard, D.I. Olive, {\it Algebras,
lattices and strings}, in: {\it Vertex operators in Mathematics and
Physics}, MSRI  Publication $\sharp 3$, Springer (1984) 51.}

\lref\sw{I. Schnakenburg, P. West, {\it Kac-Moody symmetries of IIB
supergravity}, Phys. Lett. {\bf B517} (2001) 421;
{\tt hep-th/0107181}.}

\lref\witten{E. Witten, {\it String theory dynamics in various
dimensions}, Nucl. Phys. {\bf B443} (1995) 85; {\tt hep-th/9503124}.} 

\lref\schwarz{J.H. Schwarz, {\it An SL(2,Z) multiplet of Type IIB
superstrings},  Phys. Lett. {\bf B360} (1995) 13, Erratum-ibid.
{\bf B364} (1995) 252; {\tt hep-th/9508143}.}

\lref\ait{A.C. Aitken, {\it Determinants and Matrices}, Oliver and
Boyd (1954).}  

\lref\cs{J.H. Conway, N.J.A. Sloane, {\it Sphere packings,
lattices and groups}, Springer (1999) [3rd ed.].}

\lref\no{H. Nicolai, D.I. Olive, {\it The principal SO(1,2)
subalgebra  of a hyperbolic Kac-Moody algebra}, Lett. 
Math. Phys. {\bf 58} (2001) 141; {\tt hep-th/0107146}.} 

\lref\nahm{W. Nahm, {\it Supersymmetries and their representations},
Nucl. Phys {\bf B135} (1978) 149.} 

\lref\kac{V.G. Kac, {\it Infinite dimensional Lie algebras}, Cambridge
University Press (1990) [3rd ed.].}

\lref\ruuska{V. Ruuska, {\it On purely hyperbolic Kac-Moody algebras},
in: {\it Topological and Geometrical Methods in Field
Theory}, eds.  J. Mickelsson and O. Pekonen, World Scientific
(1992) 359.}

\lref\hughes{J.W.B. Hughes, {\it Principal three dimensional
subalgebras of Kac-Moody algebras}, in: {\it Infinite dimensional Lie
algebras and their applications}, ed. S. Kass,  World Scientific 
(1988) 84.}


\Title{\vbox{\baselineskip12pt
\hbox{hep-th/0205068}
\hbox{KCL-MTH-02-10}
\hbox{SWAT-2002/337}}}
{\vbox{\centerline{A class of Lorentzian Kac-Moody algebras}}}
\smallskip
\centerline{Matthias R.\ Gaberdiel\footnote{$^\star$}{{\tt
e-mail: mrg@mth.kcl.ac.uk}}$^{,a}$,
David I.\ Olive\footnote{$^\dagger$}{{\tt
e-mail: D.I.Olive@swansea.ac.uk}}$^{,b}$ and
Peter C.\ West\footnote{$^{\ddagger}$}{{\tt
e-mail: pwest@mth.kcl.ac.uk}}$^{,a}$}
\bigskip
\centerline{\it $^a$Department of Mathematics, King's College London}
\centerline{\it Strand, London WC2R 2LS, U.K.}
\smallskip
\centerline{\it $^b$Department of Physics, University of Wales Swansea}
\centerline{\it Singleton Park, Swansea SA2 8PP, U.K.}
\vskip2cm
\centerline{\bf Abstract}
\bigskip
\noindent
We consider a natural generalisation of the class of hyperbolic
Kac-Moody algebras. We describe in detail the conditions under which
these algebras are Lorentzian. We also construct their fundamental
weights, and analyse whether they possess a real principal so(1,2)
subalgebra. Our class of algebras include the Lorentzian Kac-Moody
algebras that have recently been proposed as symmetries of
M-theory and the closed bosonic string.

\Date{May 2002}

\newsec{Introduction}

Since the realisation in the 1930's that the nuclear forces possessed
an isotopic spin symmetry, finite dimensional Lie algebras have played
an increasingly crucial role in our understanding of the fundamental
laws of nature. In particular, we now believe that three of the four
forces of nature are determined by local gauge symmetries with finite
dimensional Lie algebras. The discovery of (infinite dimensional)
Kac-Moody algebras in the late 1960s considerably enlarged the class
of Lie algebras beyond that previously considered, and a subset of
these, affine algebras, have played an important role in string theory
and conformal field theory (for a review see for example
\refs{\go}). However, until recently, no significant physical role has
been found for more general Kac-Moody algebras, namely those still
characterised by a symmetrisable Cartan matrix of finite size.

During the last few years it has become apparent that type II
superstring theory has a (non-perturbative) description in eleven
dimensional space-time in terms of M-theory \refs{\witten}. Very
little is known about the latter, but it would seem reasonable to
suppose, given previous developments, that M-theory possesses a very
large symmetry algebra. More recent work has tried to identify what
some of this symmetry could be, and it has been conjectured that it
includes a rank eleven Lorentzian Kac-Moody symmetry denoted $e_{11}$
\refs{\pwone,\pwtwo}.  Indeed, substantial fragments of this symmetry
have been found in all the maximal supergravity theories in ten and
eleven dimensions \refs{\pwtwo,\sw}. The rank eleven nature of this
symmetry can be seen to be a consequence of the bosonic field content
of the maximal supergravity theories and is related to Nahm's theorem
that supergravity theories only exist in space-times with up to eleven
dimensions \refs{\nahm}.

Simple Lie algebras of finite dimensional or affine type are well
studied and fully classified, being recognisable in terms of finite,
connected Dynkin diagrams (representing their Cartan matrices), said
to be of finite or affine type respectively.  Despite a considerable
literature on the other Kac-Moody algebras \refs{\kac}, knowledge of
their properties is much less complete. Indeed, apart from a few
cases, even the multiplicities of the various root spaces are
unknown. It is possible that for many purposes the class of all
Kac-Moody algebras may be too large and that the study of a
well-motivated subclass may be more rewarding.

One extra class of Kac-Moody algebras that has been studied in some
detail are those known as `hyperbolic'. The Dynkin diagram of a
hyperbolic Kac-Moody algebra is a {\it connected diagram such that
deletion of any one node leaves a (possibly disconnected) set of
connected Dynkin diagrams each of which is of finite type except for
at most one of affine type}. More specifically, hyperbolic Kac-Moody
algebras correspond to hyperbolic diagrams  which are the diagrams of
this type that are not of finite or affine type. The hyperbolic
Kac-Moody algebras have been classified, possess no more than ten
nodes and a Cartan matrix that is Lorentzian, that is, nonsingular and
endowed with exactly one negative eigenvalue. Furthermore every
hyperbolic Kac-Moody algebra has a real principal so(1,2) subalgebra 
\refs{\no}. Since the rank of the symmetry underlying M-theory appears
to be eleven it cannot be described by a hyperbolic Kac-Moody
algebra.  

In this paper we consider a larger class that does include the
aforementioned $e_{11}$, as well as the proposed symmetry for the
bosonic string, $k_{27}$ \refs{\pwtwo}. The Dynkin diagrams we shall
consider are  {\it connected diagrams possessing at least {\bf one}
node whose deletion leaves a (possibly disconnected) set of diagrams,
each of which is of finite type except for at most one of affine
type}.  As was noted by Ruuska, \refs{\ruuska}, such diagrams are
automatically Lorentzian if not recognisably of finite or affine type
and include the hyperbolic ones. As we shall see, the corresponding
algebras may or may not possess a real principal so(1,2) subalgebra,
for example, $e_{11}$ does, but $k_{27}$ does not.
\medskip

In section~2 we briefly recall the relations between Dynkin diagrams,
Cartan matrices and Kac-Moody algebras, and describe more precisely
the class of algebras that will be considered in this paper.  The
advantage of this class is that it is easy to determine the simple
roots in terms of those for the reduced diagram, namely the diagram
remaining when the central vertex has been deleted. The same is true
of the fundamental weights (and hence the Weyl vector as it is their
sum) and, to a lesser extent, the determinant of the Cartan
matrix. But separate treatments must be made of the two cases that the
reduced diagram possesses no affine component, or just one.

Section~3 treats the case when the reduced diagram (\ie\ the Dynkin
diagram for which the central node has been deleted) contains only
connected components of finite type. In section~4 it is shown that the
overall Dynkin diagram is Lorentzian if and only if there is precisely
one affine component given that the reduced diagram is a mixture of
connected components of finite and affine type.  Then the expression
for the determinant of the Cartan matrix simplifies considerably and
can be used to identify a large class of unimodular even Lorentzian
Cartan matrices. We also study the conditions under which these
Lorentzian algebras may possess a real principal so(1,2) subalgebra.

In section~5 we describe in detail a special subclass of constructions
that seems to be of particular relevance in string theory. We also
discuss more specifically the algebras that are associated to even
self-dual lattices, and study the question of whether the algebras 
possess a real principal so(1,2) subalgebra. Section~6 describes
further constructions and section~7 contains some conclusions. We have  
added four appendices in which various more technical points that
are needed for our discussion are outlined.

\newsec{A special class of Kac-Moody algebras}

First let us recall the definition of a Kac-Moody algebra
\refs{\kac} in terms of a generalised Cartan matrix. Suppose 
$A_{ij}$ is a generalised Cartan matrix with $i,j=1,\ldots, r$, where
$r$ is the rank of the Kac-Moody algebra. We shall only consider
generalised Cartan matrices that are symmetric. They satisfy
$$A_{ii}=2\,,
\eqno(2.1)$$
$$A_{ij}=A_{ji}\ \  {\rm for}\  i\not= j\  {\rm are\  negative\
integers\  or \ zero}\,.
\eqno(2.2)$$
Because the off diagonal entries, (2.2) could possibly take values
$-2$, $-3$ {\it etc}, and it is not appropriate to refer to the case
in which $A$ is symmetric as being simply-laced, unless the matrix is
of finite or affine type and such values are disallowed.  The entries
of the matrix $A_{ij}$ can be encoded in terms of an unoriented graph
with $r$ nodes, whose adjacency matrix is given by
$2\delta_{ij}-A_{ij}$.\footnote{$^\star$}{The $(i,j)$ entry of the     
adjacency matrix of an unoriented graph describes the number of links
between the nodes $i$ and $j$. Our conventions here differ slightly
from those of Kac \refs{\kac}.} This graph is called the Dynkin
diagram, and it specifies the  matrix $A_{ij}$ uniquely (up to
simultaneous relabelling of the rows  and 
columns). If the diagram is disconnected, the Cartan matrix has a
block diagonal form (when labelling is ordered suitably) and the
algebra consists of commuting simple factors.

Given a generalised Cartan matrix, the Kac-Moody algebra can be
formulated in terms of a set of Chevalley generators $H_i$, $E_i$ and
$F_i$ for each $i=1,\ldots,r$. These can be identified with the
generators of the Cartan subalgebra, and the generators of the
positive and negative simple roots, respectively. The Chevalley
generators are taken to obey the Serre relations
$$[H_i, H_j]= 0\,,
\eqno(2.3)$$
$$[H_i, E_j]= A_{ij} E_j\,,
\eqno(2.4)$$
$$[H_i, F_j]= -A_{ij} F_j\,,
\eqno(2.5)$$
$$[E_i, F_j]= \delta_{ij} H_i\,,
\eqno(2.6)$$
and
$$[E_i,\ldots [E_i, E_j]\ldots ]= 0\,, \qquad
[F_i,\ldots [F_i, F_j]\ldots]= 0\,.
\eqno(2.7)$$
In equation (2.7) the number of $E_i$'s in the first equation, and the
number of $F_i$'s in the second is $1-A_{ij}$. The remaining
generators of the Kac-Moody algebra are obtained as multiple
commutators of the $E_i$'s and as multiple commutators of the
$F_i$'s, using the above Serre relations. The generalised Cartan
matrix therefore determines the Kac-Moody algebra uniquely. Although
this procedure is fairly simple in principle, explicit descriptions
for all generators of a Kac-Moody algebra are only available for a
few, rather special, algebras. 

A convenient  basis for the Kac-Moody algebra consists of one for the 
Cartan subalgebra (which has dimension $r$) consisting of the
Chevalley generators $H_i$ with $i=1,\ldots,r$,  and the step
operators for roots. The roots are eigenvectors of the Cartan
generators  (under the adjoint action), and we can therefore think of
them as vectors in an $r$-dimensional vector space. There exists a
scalar product $(\hbox{ },\hbox{ })$ on this space such that the
Cartan matrix is given in terms of the simple roots as  
$$
A_{ij} =(\alpha_i,\alpha_j)\,.\eqno(2.8)
$$
This means that, if the Cartan matrix $A$ is non-singular, its
signature and rank  is the same as that of the scalar product.  
If $A$ is singular there is an analogous but more complicated
statement. Thus (2.1) implies that all the simple roots have length
$\sqrt2$ and, in particular,  are  space-like. Their integer span is
an even lattice, known as the root lattice, denoted
$\Lambda_R(A)$. The root lattices of inequivalent Kac-Moody algebras
may or may not be equivalent; indeed, we shall encounter examples of 
inequivalent Kac-Moody algebras (that are described by inequivalent
Dynkin diagrams) with the same root lattice.

Only if the Cartan matrix $A$ is positive definite  is the associated
algebra of finite dimension. Then the diagram and Cartan matrix is
said to be of finite type. If $A$ is positive semi-definite it is
said to be of affine type. We shall mainly be interested in those
Kac-Moody algebras that are {\it Lorentzian},  namely those whose
generalised Cartan matrix  is non-singular  and possesses precisely
one negative eigenvalue. 
\medskip

We have already mentioned the fully classified subset  known as
`hyperbolic' Kac-Moody algebras. In this paper we will study a larger
class of Kac-Moody algebras which includes the finite, affine and
hyperbolic types and is automatically  Lorentzian if neither of finite
nor affine type. These correspond to a {\it Dynkin diagram
possessing at least {\bf one} node  whose deletion yields a  diagram
whose connected components are of finite type except for at most one
of affine type}. 

We shall call the overall Dynkin diagram $C$, and the  selected node,
whose deletion yields the reduced diagram $C_{R}$, the ``central''
node. For many examples the central node is not uniquely determined
by the property that $C_R$ has only connected components of finite and
affine type, and what we shall do in the following will apply to every
admissible choice for the central node. Unlike the overall $C$, the
reduced diagram $C_{R}$ need not be connected. If it is disconnected
denote the $n$ connected components $C_1, C_2\dots C_n$. The Cartan
matrix of $C_{R}$ is obtained from that for $C$ simply by deleting the
row and column corresponding to the central node. 
Then the  overall Dynkin diagram  can be re-constructed from the
reduced Dynkin  diagram and the central node, denoted $c$, by adding
those edges linking the latter to each node of $C_{R}$. The
number of links of the central node $c$ to the $i$'th node is 
$$\eta_i=-A_{ci}\,,\eqno(2.9)$$
namely the entries in the row and column of the Cartan matrix of $C$
whose deletion was just mentioned. 

Note that if the connected components of $C_{R}$   are all of
finite or affine type, the off-diagonal elements in its Cartan matrix 
only take the values $0$ or $-1$ (for convenience  $A_1^{(1)}$ in
standard notation is excluded). This limitation does not apply to the
values of $\eta_i$ which could be any integer
$0,1,2\dots$. Schematically, the Dynkin diagram $C$ has the
structure:
\ifig\dynkinpic{The Dynkin diagram $C$ of the Kac-Moody
algebra $\g$ for the case
$n=3$.}{\epsfxsize2.5in\hskip-.5cm\epsfbox{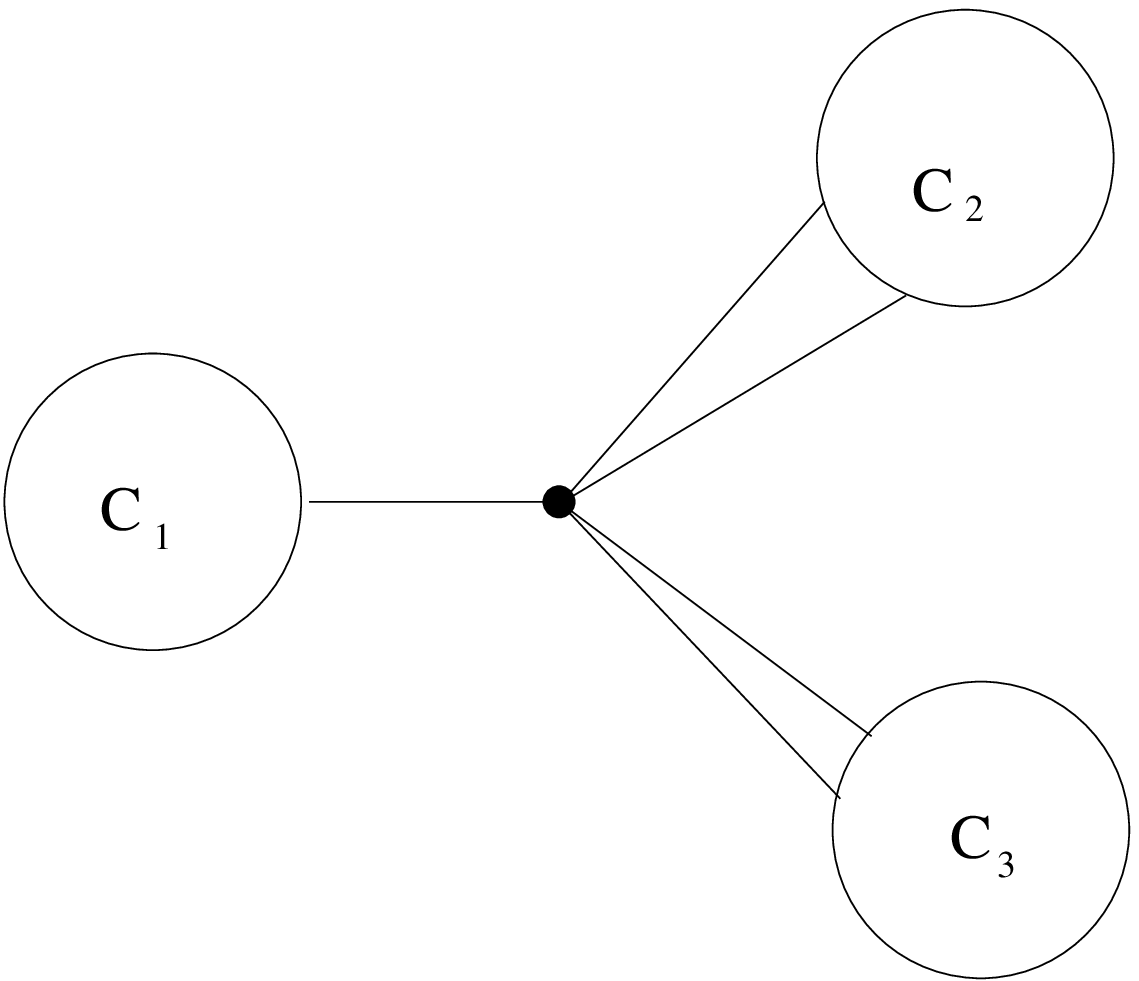}} 
Notice that the diagram $C$ need not be a tree diagram and indeed may
possess any number of loops. 

If the Cartan matrix $A$ is non-singular it is possible to define
fundamental weights $\lambda_1,\lambda_2, \dots \lambda_r$ reciprocal
to the simple roots 
$$\lambda_i=\sum_{j=1}^r\, (A^{-1})_{ij}\alpha_j\,,
\quad \lambda_i.\alpha_j=\delta_{ij}\,.\eqno(2.10)$$ 
For hyperbolic algebras these  all lie inside or on the same
light-cone (so that the entries $(A^{-1})_{ij}=\lambda_i.\lambda_j$
are all negative or zero), but for Lorentzian algebras that are not
hyperbolic some fundamental weights must be space-like. 

The integer span of the fundamental weights forms a lattice known as the
weight lattice $\Lambda_W(A)$. It is reciprocal to the root lattice
$\Lambda_R(A)$ yet contains it. Hence it is possible to consider the
quotient which forms a  finite abelian group $\Z(A)$ containing
$|\Z(A)|$ elements  
$$\Lambda_W(A)/\Lambda_R(A)=\Z(A)\,.\eqno(2.11a)$$
Of course this group is encoded in the Cartan matrix whose
determinant, up to a sign, equals the number of elements it contains 
$$ \det\,A=\pm|\Z(A)|\,.\eqno(2.11b)$$
If $A$ is of finite type it is associated with a finite dimensional 
semi-simple Lie algebra $\g$ which can be exponentiated uniquely to a
simply connected Lie group $G$ whose centre is the finite abelian
group $\Z(A)$. 

Given the fundamental weights, the Weyl vector $\rho$ also exists
$$\rho=\sum_{i=1}^r\lambda_i=\sum_{i,j=1}^r(A^{-1})_{ij}\,
\alpha_j\,, \eqno(2.12a)$$    
satisfying the fundamental property
$$\rho.\alpha_i=1\,,\quad i=1,2\dots r\,.\eqno(2.12b)$$
A critical question for the existence of a principal so(1,2)
subalgebra with desirable reality properties is whether the
coefficients $\sum_{i=1}^r\,(A^{-1})_{ij}$ in (2.12a)  are all of the
same sign or not. More precisely, by a real principal so(1,2)
subalgebra is meant one that has standard hermiticity properties given 
the hermiticity properties of the Kac-Moody algebra. This has
desirable consequences for unitarity, and more details and
explanations are given in appendix~A. It leads to more stringent
conditions than those discussed by Hughes \refs{\hughes}. We shall
only consider real principal so(1,2) subalgebras in this paper, and we
shall therefore drop from now on the qualifier `real'.
In the finite case the coefficients $\sum_{i=1}^r\,(A^{-1})_{ij}$
are all positive, while they are all negative in the hyperbolic case
and in the Lorentzian case  mixed signs are possible. We shall show
that, within the class defined above, one of these signs at least is
negative.

\newsec{When the reduced diagram is of finite type}

In this case  we shall construct linearly independent simple roots for 
the overall Dynkin diagram $C$ in terms of those for the reduced
diagram $C_{R}$ as well as doing the same for the fundamental
weights when that is possible. A formula for the determinant of the
Cartan matrix of $C$ will likewise be found. 

The $r-1$ simple roots for $C_{R}$, 
$\alpha_1,\alpha_2,\dots \alpha_{r-1}$, are linearly independent  
and span a Euclidean space of dimension $r-1$ since the reduced
diagram $C_{R}$ is of finite type. They will suffice for the
corresponding nodes of the overall Dynkin diagram $C$  once they are
augmented by the simple root for the central node 
$$\alpha_c=-\nu+x\,,\quad\hbox{where}\quad
\nu=\sum_{i=1}^{r-1}\eta_i\lambda_i=-\sum_{i=1}^{r-1}A_{ci}\lambda_i\,.
\eqno(3.1)$$ 
Here $\lambda_i$ are the fundamental weights (2.10) for the reduced
diagram (which is assumed to have a non-singular Cartan matrix) and
lie in the space spanned by the simple roots, while $x$ is a vector
orthogonal to that space. Evidently this guarantees
$\alpha_i.\alpha_c=A_{ic}$ leaving only the condition
$$2=A_{cc}=\nu^2+x^2\eqno(3.2)$$
which determines the sign of $x^2$. If $C_{R}$ is of finite
type  (whether disconnected or not), its simple roots span a Euclidean
space. Thus the simple roots of $C$ span a space that is Euclidean or 
Lorentzian according as $x^2$ is positive or negative. Likewise if
$x^2$ vanishes the simple roots span a space with a positive
semi-definite metric and so constitute an affine root system. Since
Dynkin diagrams of finite or affine type are fully classified they are
recognisable as such. Hence if $C$ is of neither of these types it
must be Lorentzian, as claimed earlier. It is at this stage that the
connection (2.8) between the scalar product and the Cartan matrix is
exploited. 

The $r$ fundamental weights for the overall diagram $C$ will be
denoted $\ell_c,\ell_1,\ell_2,\dots \ell_{r-1}$ in order to
distinguish them from the $(r-1)$ fundamental weights 
$\lambda_1,\dots, \lambda_{r-1}$ for the reduced diagram. 
They are related to each other by
$$\ell_i=\lambda_i+{\nu.\lambda_i\over x^2}x\,, \quad 
\ell_c={1\over x^2}x\,, \eqno(3.3)$$ 
providing $x^2$ does not vanish, that is the overall diagram $C$ is
not affine. This accords with the fact that the definition (2.10)
fails only in this case. The overall Weyl vector is then
$$R\equiv\sum_{j=c}^{r-1}\ell_j=\rho+{(1+\nu.\rho)\over x^2}x\,,
\eqno(3.4)$$
where $\rho=\sum_{j=1}^{r-1}\lambda_j$ is the Weyl vector for the
reduced diagram, (the same as the sum of the Weyl vectors for each
connected component of $C_{R}$ if it is disconnected). Notice
that 
$$R.\ell_c={(1+\nu.\rho)\over x^2}\,,\eqno(3.5)$$
and hence has the same sign as $x^2$ given that $\nu.\rho$ is positive
when  $C_{R}$ is of finite type (as all quantities
$\lambda_i.\lambda_j$ are). Thus at least one of the coefficients in 
the expansion of the Weyl vector $R$ in terms of simple roots is
negative when $C$ is Lorentzian. 

An instructive example is provided by choosing the linking
coefficients $\eta_i$ to  equal each other, taking the value 
$\bar \eta$ say, so that the central node is  linked by precisely
$\bar\eta$ edges to each other node. Then $\nu$ in (3.1) equals
$\bar\eta$ times the Weyl vector $\rho$ and,  by (3.2),
$x^2=2-\bar\eta^2\rho^2$. Hence equations (3.3) and (3.4) lead to 
$$R.\ell_c={\bar\eta\rho^2+1\over x^2}\,,\qquad 
R.\ell_i={(2+\bar \eta)\rho.\lambda_i\over x^2}\,,
\eqno(3.6)$$
which are all negative if $x^2$ is, that is if $C$ is Lorentzian. 
In particular, this therefore means that there are infinitely many
Lorentzian algebras with a principal so(1,2) subalgebra that can be
obtained by this construction since we can choose $\g$ to be any 
finite dimensional semi-simple Lie algebra. The simplest example is
obtained by considering $\bar\eta=1$, and taking $\g={\rm su}(m)$ for
$m\geq 4$. The Dynkin diagram for the case  $m=9$ is shown below.
\ifig\examplepic{The Dynkin diagram of the Kac-Moody algebra $\g$
obtained by taking $\g_1={\rm su}(9)$ and $\nu= \rho$.}
{\epsfxsize4.0in\hskip.2cm\epsfbox{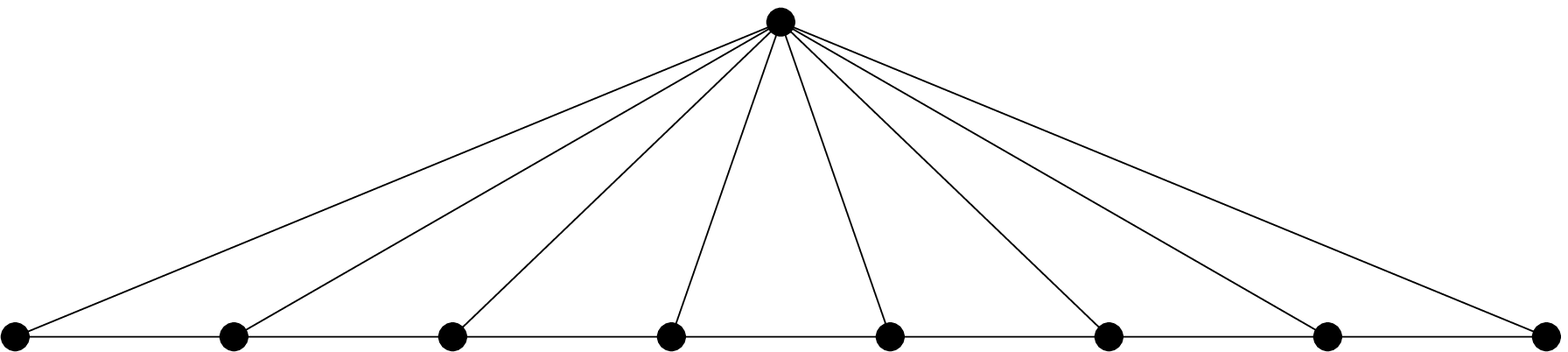}}

The determinant of the  Cartan matrix $A$ for the overall diagram $C$
is related to the Cartan matrix $B$ of the reduced diagram
$C_{R}$ (obtained from $A$ by deleting the row and column
corresponding to the central node) by 
$$\hbox{det}\,A=x^2\,\hbox{det}\,B=(2-\nu^2)\, \hbox{det}\,B\,,
\eqno(3.7)$$ 
or, using (3.1) and the fact that $\lambda_i.\lambda_j=(B^{-1})_{ij}$ 
$$\hbox{det}\,A=\left(2-\sum_{i,j=1}^{r-1}
\eta_i(B^{-1})_{ij}\eta_j\right)
\hbox{det}\,B\,, \eqno(3.8)$$
or remembering that the adjugate of a matrix (the matrix of
cofactors) equals the inverse matrix times the determinant
$$\hbox{det}\,A=2\,\hbox{det}\,B-\sum_{i,j=1}^{r-1}
\eta_i(\hbox{adj}\,B)_{ij}\eta_j\,.\eqno(3.9)$$ 
This final version makes it clear that the result is indeed an integer
even though the sign is unclear. Equation (3.9) also has the virtue
that it makes good sense even when $B$ is singular. Indeed it
simplifies considerably as the first term on the right hand side drops
out. We shall return to this point in the next section. 

Equation (3.7) can be proven directly by using (2.8) to factorise
$\det\, A$ into products of determinants of matrices made of the
components of the simple roots of $C$. The crucial point is that the
simple roots for nodes of $C_{R}$ have no component in the
direction of $x$ and this makes it trivial to evaluate the factored
determinants. Alternatively (3.9) is just an application of Cauchy's
expansion of bordered determinants \refs{\ait}. 

Only now is account taken of the fact that the reduced diagram
$C_{R}$ may be disconnected with connected  components
$C_1,C_2,\dots C_n$ as depicted in Fig 1. The consequence is that
after a suitable reordering of rows and columns the Cartan matrix $B$
is block diagonal 
$$B=\hbox{diag}(B_1,B_2,B_3\dots, B_n)\,, \eqno(3.10)$$
where $B_{\beta}$ is the Cartan matrix of the Dynkin
diagram for $C_{\beta}$. It is convenient to denote
$\Delta_{\beta}=\hbox{det}\,B_{\beta}$. Then (3.9) reads
$$\hbox{det}\,A=\Delta_1\Delta_2\dots\Delta_n
\left(2-\sum_{\beta=1}^n\left\{{
\sum_{i,j\in C_{\beta}}\eta_i
\left({\hbox{adj}\,B_{\beta}}\right)_{ij}\eta_j)\over\Delta_{\beta}}
\right\}\right)\,.\eqno(3.11)$$
This identity provides an efficient tool for evaluating determinants
of Cartan matrices iteratively as will be illustrated in the case
where the central node is linked to each disjoint component
$C_{\beta}$ by a single edge attached to a distinguished node of
$C_{\beta}$ that is denoted by $*$. If $B_{\beta}^*$ denotes the
Cartan matrix obtained from $B_{\beta}$ by deleting the row and column
corresponding to the node $*$ (and is automatically of finite type if
$B_{\beta}$ is), and  $\Delta_{\beta}^*=\hbox{det}\,B_{\beta}^*$,
$$\hbox{det}\,A=\Delta_1\Delta_2\dots \Delta_n
\left(2-\sum_{\beta=1}^n{\Delta_{\beta}^*\over\Delta_{\beta}}\right)\,.
\eqno(3.12)$$

Notice that when $C_n$ contains no nodes it can be deemed  to be an
empty diagram so that the reduced diagram $C_{R}$ contains
only $n-1$ connected components. The result  (3.12) ought to reflect
this fact and it does so if it is understood that  $\Delta_n=1$ and
$\Delta_n^*=0$ for an empty diagram.  

Let us now use (3.12) to determine a few determinants explicitly. For
the case of $su(N)$, det$\,A(su(N))\equiv\Delta(su(N))$ is evaluated
by considering the $a_{N-1}=su(N)$ Dynkin diagram and selecting as the
central node one of the two end nodes so that the reduced diagram is
connected. Then $\Delta_1=\Delta(su(N-1))$,
$\Delta_1^*=\Delta(su(N-2))$ and (3.12) reduces to 
$$\Delta(su(N))=2\Delta(su(N-1))-\Delta(su(N-2))\,.\eqno(3.13)$$
This is a simple recurrence relation whose general solution is
$\Delta(su(N))=AN+B$. The constants $A$ and $B$ are determined as $1$
and $0$ respectively  by the comments above concerning empty diagrams
which imply $\Delta(su(1))=1$ and $\Delta(su(0))=0$, yielding the
familiar result 
$$\Delta(su(N))=\hbox{det}\, A(su(N))=N\,.\eqno(3.14)$$

A similar argument applies to the Dynkin diagram of the Lie
algebra $d_N=so(2N)$ by taking as central node one of the two spinor
tips. Again the reduced diagram is connected but this time 
$\Delta_1=\hbox{det}\,A(su(N))=N$ and 
$\Delta_1^*=\hbox{det}\,A(su(2))\,\hbox{det}\,A(su(N-2))=2(N-2)$ by
the  results for $su(N)$. Then (3.12) yields another familiar result
$$\Delta(so(2N))=\hbox{det}\,A(so(2N))=4\,.\eqno(3.15)$$ 

More interesting is the Dynkin diagram for $e_N$. Selecting as central
node the tip of the shortest leg yields
$\Delta_1=\hbox{det}\,A(su(N))=N$ and
$\Delta_1^*=\hbox{det}\,A(su(3))\,\hbox{det}\,A(su(N-3))=3(N-3)$, and
so (3.12) gives 
$$\hbox{det}\,A(e_N)=9-N\,.\eqno(3.16)$$
This indicates that $e_N$ is Lorentzian if $N\geq10$, as indeed it is 
by the preceding discussion. Notice also that $e_{10}$, which is
hyperbolic, has a Cartan matrix with determinant $-1$ so that its root
lattice $\Lambda_R(e_{10})$ is an even, unimodular Lorentzian lattice,
a somewhat rare object. In the next section we shall find many more
Cartan matrices for such lattices. As explained below $e_{10}$ can be
thought of as what is called an overextension of the
finite dimensional Lie algebra $e_8$. Likewise $e_{11}$ whose Cartan
matrix has determinant $-2$ can be viewed as a very extended version
of $e_8$. 

\newsec{When the connected components of the reduced diagram are
either finite or affine type} 

Suppose that $p$ is the number of connected components of
$C_{R}$ that are of affine type. Thus $p$ factors
$\Delta_{\beta}=\hbox{det}\,B_{\beta}$ vanish, so that, taking account
of cancellations it appears from (3.11) that $\hbox{det}\,A$ has a
$(p-1)$-fold zero. In fact $A$ does have corank $(p-1)$ so that only
if $p=1$ is $C$ a Lorentzian diagram. Otherwise it is neither
Lorentzian nor affine as its  Cartan matrix $A$ has one negative
eigenvalue and a $(p-1)$-fold zero eigenvalue. 

This is established by displaying a set of simple roots whose scalar
products yield the Cartan matrix $A$ whilst spanning a space of
dimension $(r-p+1)$ equipped with a Lorentzian scalar product. 

First, simple roots are assigned to the reduced diagram, component by
component. For each  component $C_{\beta}$ assign the simple roots
$\alpha_i$, $i\in C_{\beta}$. If $C_{\beta}$ is of finite type these
are linearly independent whilst if it is of affine type these are
linearly dependent,  
$$\sum_{i\in C_{\beta}}\,n_i\alpha_i=0\,,$$
where the positive integers $n_i$ are the Kac labels for the affine
diagram $C_{\beta}$. Then the simple roots assigned to the overall
diagram $C$ are, in terms of these, 
$$a_i=\alpha_i+\eta_i\,k=\alpha_i-A_{ci}\,k\,, \quad 
i\in C_{R}\,, \eqno(4.1)$$
$$a_c=-(k+\bar k)\,,\quad\hbox{where}\quad k^2=\bar k^2=0\,,\quad  
k.\bar k=1\,.\eqno(4.2)$$
The vectors $k$ and $\bar{k}$ can be thought to lie in the even
self-dual Lorentzian lattice $\II^{1,1}$ whose structure is described
in appendix~B. 

The scalar products of the $a_i$ realise the overall Cartan matrix $A$  
with $r$ rows and columns yet the roots manifestly span a space  of
dimension $r-p+1$. Since for each of the $p$ affine components of the
reduced diagram $C_{\beta}$ 
$$\sum_{i\in C_{\beta}}n_ia_i=
\left(\sum_{i\in C_{\beta}}n_i\eta_i\right)\, k\,,\eqno(4.3)$$ 
elimination of $k$ yields $p-1$ linear relations amongst these simple
roots. 

Let us henceforth concentrate on the case that $A$ is Lorentzian so
that $p=1$. Let the affine component of the reduced diagram be $C_1$
so that $C_2,  C_3,\dots C_n$ are all of finite type. Then
$\Delta_1=\hbox{det}\,B_1$ vanishes and the right hand side of
equation (3.11) simplifies as $n$ of the $(n+1)$ terms vanish, leaving 
only the $\beta=1$ term in the sum 
$$\hbox{det}\,A=-\Delta_2\dots \Delta_n\sum_{i,j\in C_1}
\eta_i(\hbox{adj}\,B_1)_{ij}\eta_j\,.\eqno(4.4)$$

Now $C_1$ is a  connected simply-laced affine diagram and so it has to
have the form of an affine Dynkin diagram for a simple, simply-laced
affine Kac-Moody algebra $\g^{(1)}$, say. So $B_1=B(\g^{(1)})$. 
Then its adjugate matrix has the form
$$(\hbox{adj}\,B_1)_{ij}=|\Z(G)|\,n_i\,n_j\,,\eqno(4.5)$$ 
where again the integers $n_i$ are the Kac labels for
$\g^{(1)}$. They constitute the unique null vector of $B_1$,
$(B_1)_{ij}n_j=0$.  But, 
by definition, its adjugate matrix satisfies
$(B_1)_{ij}(\hbox{adj}\,B_1)_{jk}=\delta_{ik}\hbox{det}\,B_1=0$. Hence
each column of $\hbox{adj}\,B_1$ is proportional to the null
vector. The structure above then follows from the fact that
$\hbox{adj}\,B_1$, like $B_1$ is symmetric. The normalisation follows
by specialising the suffices $i$ and $j$ to the value $0$, denoting
the affine node, and remembering that $n_0=1$ while
$(\hbox{adj}\,B_1)_{00}$ is the determinant of the Cartan matrix for
$\g$ and hence equals $|\Z(A(\g))|$ by (2.11b) and comments
thereafter.  

Hence $\hbox{det}\,A$ further simplifies 
$$\hbox{det}\,A=-\Delta_2\Delta_3\dots \Delta_n\,|\Z(G)|\,
\left(\sum_{i\in C_1}n_i\eta_i\right)^2\,,\eqno(4.6)$$ 
and $\hbox{det}\,A$ is explicitly negative, being expressed as minus a
product of positive integers. Notice the remarkable fact that any
dependence on the quantities $\eta_i=-A_{ci}$, $i\not\in C_1$ has
disappeared. 

The root lattice of any of the diagrams under consideration is an
even, integral Lorentzian lattice and, by (2.11) it is self-dual, or
self-reciprocal if and only if $\hbox{det}\,A$ equals $-1$. 
By (4.6) this is  so only if each factor on the right hand side, being
an integer, actually  equals unity. The only simply-laced  connected
diagram of finite type with unimodular Cartan matrix is, by the
results of the preceding section, the $e_8$ Dynkin diagram. So $C_2$,
$C_3,\dots C_n$ must each be of this type. So also must $\g$ be $e_8$
so that $C_1$ must be an affine $e_8$ diagram, or equivalently, an
$e_9$ diagram. Finally the factor $\sum_{i\in C_1}n_i\eta_i$ must
equal unity. Thus all $\eta_i$ here vanish,  except for just one that
equals unity and must correspond to  the node of $C_1$ for which
the Kac index equals unity. This is the affine node (or one related to
it by a diagram symmetry). Thus the central node of $C$ is linked to
$C_1$ in effect only via the affine node. The dimension of the even,
Lorentzian self-dual lattice has therefore to be $8n+2$, in accord
with the fact that it is only in these dimensions that such lattices
exist. They are denoted $\II^{8n+1,1}$ and are unique. (For a brief
description of these lattices see appendix~B.) Nevertheless,
notice that, because of the arbitrariness in the quantities
$\eta_i,\, i\not\in C_1$, there are very many Cartan matrices (and
therefore many inequivalent Kac-Moody algebras) that give rise to each
of these when $n>1$. If $n=1$ this procedure yields only one Cartan
matrix whose root lattice is $\II^{9,1}$ and that is the $e_{10}$
Cartan matrix previously mentioned as an over-extension of the $e_8$
Cartan matrix.

The fundamental weights for the overall diagram $C$ are determined in
terms of the fundamental weights associated with the reduced diagram
as 
$$\ell_{\beta}=\lambda_{\beta}\,,\quad \beta\in C_2\,,\eqno(4.7a)$$
$$\ell_c=-k\,, \eqno(4.7b)$$
$$\ell_i=\lambda_i-{n_i\over\eta}(k-\bar k+\nu)\,,\quad 
i\in C_1\,.\eqno(4.7c)$$
To simplify notation all components of $C_2$ of finite type are
included in $C_2$ which is no longer taken to be
connected, and $\lambda_{\beta}$ are the fundamental weights of
$C_2$. As already mentioned $C_1$ has to be the extended Dynkin
diagram of a finite dimensional simply-laced simple Lie algebra, $\g$,
or equivalently the Dynkin diagram for the untwisted affine Kac-Moody 
algebra $\g^{(1)}$. $\lambda_1,\lambda_2,\dots \lambda_{r_1}$  are the
fundamental weights of $\g$ and $\lambda_0=0$. $\nu$ records the
linkage of the central node to the nodes of $C_{R}$ 
$$\nu=\nu(\g)+\nu(C_2)\,, \quad \hbox{ where }\quad
\nu(\g)=\sum_{i=1}^{r(\g)} \eta_i\lambda_i \quad
\hbox{and}\quad \nu(C_2)=\sum_{\beta\in C_2}
\eta_{\beta}\lambda_{\beta}\,,\eqno(4.8)$$
and $\eta$ is the quantity that already appeared in the determinant
formula (4.6), namely 
$$\eta=\sum_{i\in C_1} n_i\eta_i\,.\eqno(4.9)$$
Notice that $\eta_0=-A_{c0}$ contributes to $\eta$ but not to
$\nu$. It is easy to check that the weights (4.7) do satisfy (2.10),
given (4.1) and (4.2). 

So the Weyl vector $R$ for the overall diagram $C$, being the sum of
the fundamental weights, is
$$R=\rho(C_2)+\rho(\g)-{h(\g)\over\eta}(k-\bar k+\nu)-k\,, 
\eqno(4.10)$$  
where $\rho(\g)=\sum_{j=1}^{r(\g)}\lambda_j$ is the Weyl vector for
$\g$, $h(\g)=\sum_{j=0}^{r(\g)}n_j$ is the Coxeter number of $\g$ and 
$\rho(C_2)$ is the Weyl vector for $C_2$. Notice immediately that 
$$R.\ell_c=-{h(\g)\over\eta}<0\,.\eqno(4.11)$$
This is very similar to what happened in the previous section,
equation (3.5), and means that if there is a principal three
dimensional subalgebra it must be so(1,2) rather than so(3). Also
$$R^2=\left[\rho(C_2)-{h(\g)\over\eta}\nu(C_2)\right]^2+
\left[\rho(\g)-{h(\g)\over\eta}\nu(\g)\right]^2
-{2h(\g)(h(\g)+\eta)\over\eta^2}\,.\eqno(4.12)$$ 
The only negative term is the last and it depends on $C_1$ and its
linkage to the central node and not at all on $C_2$. 

Let us consider in turn two possibilities for the linkage between the
central node and the affine component $C_1$. First suppose that the
only link is to the affine node so $\eta_i=\delta_{i0}, i\in C_1$. 
Then $\eta=1$ and $\nu(\g)$ vanishes so that (4.12) reduces to 
$$R^2=\rho(\g)^2-2h(\g)(h(\g)+1)+[\rho(C_2)-h(\g)\nu(C_2)]^2\,.
\eqno(4.13)$$
This can be simplified by the Freudenthal-de Vries strange formula
applied to $\g$, 
$$\rho(\g)^2={h(\g)(h(\g)+1)r(\g)\over12}\,, \eqno(4.14)$$
to yield
$$R^2={h(\g)\, (h(\g)+1)\, (r(\g)-24)\over12}
+[\rho(C_2)-h(\g)\nu(C_2)]^2\,.
\eqno(4.15)$$ 
This cannot be negative unless $\g$ has rank $r(\g)$ less than
$24$. Since this is another necessary condition for the presence of a
principal so(1,2) subalgebra, it means that there is only a finite
number of possibilities for $\g$ in this situation. There are also
constraints on $C_2$, as will be discussed below in section~4.1.

A particularly interesting case is when $C_2$ is empty and the
Lorentzian algebra with Dynkin diagram $C$ is said to be an 
``overextension'' of $\g$. Then the condition $R^2<0$ reduces to
$r(\g)<24$, as noted some time ago \refs{\go}. Thus (4.15) can be
regarded as a generalisation of this result. 

Consider now the remaining possibilities for a single link between the
central node and the affine diagram $C_1$ so $\eta_i=\delta_{i*}$
where $i,*\in C_1$ and the Kac label of the node $*$ is $n_*\geq2$. As 
all nodes of the affine $su(N)$ diagrams have unit Kac label the only 
possibilities for the finite dimensional Lie algebra $\g$ are
$e_6,e_7,e_8$ and $d_n$. For any choice of node in $e_6, e_7$ and
$e_8$ a calculation reveals that the sum of the last two terms in
(4.12) is negative so that the squared length of the overall Weyl
vector could be negative for some choices of $C_2$. If $\g$ is $d_n$
the only possibility is $n_*=2$. Then deletion of node $*$ from $C_1$
which is the affine $d_n$ diagram yields a Dynkin diagram of type
$d_p\oplus d_q$, where $p+q=n$ and $p,q\geq 2$. The sum of the last
two terms of (4.12) is 
$${(n-1)\over12}\left(n(n-26)+3(p-q)^2\right)\,.\eqno(4.16)$$
This is negative for a finite number of choices for $p,q$  all
entailing $n=p+q<26$.  

The conclusion is that whenever there is a single link between the
central node and $C_1$, the number of possibilities for the affine 
diagram $C_1$ and its linked node is finite when (4.12) is
negative. It will be shown below that likewise the number of
possibilities for $C_2$ is also finite.  

Now we look at the choice of the $\eta_j$ that seems most likely to
produce a principal so(1,2) by minimising $R^2$, (4.12), given
$\g$. First force the middle term on the right hand side of (4.12) to 
vanish by taking $\eta\rho(\g)=h(\g)\nu(\g)$. The necessary and
sufficient condition for this is that all $\eta_j$, $j\in C_1$ be
equal, to $\eta_0$, say. The first term then vanishes 
if and only if $\eta_j$, $j\in C_2$ all equal $\eta_0$. In this case  
$$
R.\ell_{\beta}=0\,,\quad\beta\in C_2\,, \quad
R.\ell_c=-{1\over\eta_0}\,, \; \hbox{ and }\; 
R.\ell_j=-{n_j\over h(\g)\eta_0}
\left(1+{2\over\eta_0}\right)<0\,. \eqno(4.17)$$ 
So $R.\ell_i<0$, $j\in C_{R}$ only if $C_2$ is empty. In that
case the Lorentzian algebra corresponding to $C$ always has a
principal so(1,2) subalgebra, whatever $\g$.

\subsec{Constraints on $C_2$}

As we have seen above, if there is a single link between the central
node and $C_1$, the number of possibilities for the affine diagram
$C_1$ and its linked node is finite if (4.12) is negative. We want
to show now that for each of the finitely many choices for $C_1$,
there are only finitely many choices for $C_2$ that make (4.12)
negative. In particular, this shows that within this class of
algebras, the rank of the algebras that possess a principal so(1,2)
subalgebra is bounded from above.

\noindent Given $\g$, the condition that (4.12) is negative is simply
that  
$$
\left[ \rho(C_2) - h_0\, \nu(C_2) \right]^2 \leq M_0(\g) \,,
\eqno(4.18)
$$
where $h_0={h(\g)\over \eta}$, and 
$M_0(\g)={2 h(\g) (h(\g)+\eta) \over \eta^2}-
\left(\rho(\g) - h_0 \, \nu(\g)\right)^2$ only depend on $\g$. By
considering the different possibilities for the algebra $\g$ it is
easy to see that $h_0\geq 2$ for each simply-laced Lie algebra. 

If $C_2$ is not connected, we can split the right hand side of (4.18)
into a sum over the simple components. For each simple component, the
left hand side of (4.18) is strictly positive since $h_0\geq 2$. This 
suffices to show that we can only have finitely many simple components
in $C_2$. It therefore remains to show that the rank of each simple
component must be bounded. This will be done separately for $a_r$ and
$d_r$. In the following we shall write $\rho=\rho(C_2)$,
$\nu=\nu(C_2)$. 
\smallskip

\noindent {\bf The case of $a_r$}

\noindent Let us write the vector $\nu$ in the orthogonal basis of
appendix~C, \ie\
$$
\nu = \sum_{j=1}^{r+1} l_j\, e_j\,, \eqno(4.19)
$$
where $l_j$ depends on $\nu$. Given the formula for the fundamental
weights (C.7), it now follows that $l_{j+1}-l_j = -\eta_{j}$.
Thus if we write $(\rho - h_0\; \nu)$ in the same basis,
$$
(\rho - h_0\; \nu) = \sum_{j=1}^{r+1} m_j \, e_j \,,
\eqno(4.20)
$$
then $m_{j+1} - m_j = \eta_{j} \, h_0 - 1$.
Since $h_0\geq 2$ and $\eta_{j}\in\Nop_0$, at least every other
$m_j$ is in modulus bigger or equal to $1/2$, and therefore
$$
(\rho - h_0\; \nu)^2 \geq {r\over 8} \,.
\eqno(4.21)
$$
Because of (4.18) it is then immediate that the rank of $C_2$ must be
bounded. It is also obvious from the above argument that only finitely
many choices for $\eta_\beta$, $\beta\in C_2$ will respect the bound
(4.18). 
\smallskip

\noindent {\bf The case of $d_r$}

\noindent Let us first consider the case when $\nu$ is not a spinor
weight. Then, given the formula for the fundamental weights (C.13)
and the Weyl vector (C.14) it follows that
$$
(\rho - h_0\; \nu ) =
\sum_{i=1}^{r} \left( r - i - h_0\; l_i \right) e_{i}\,, \eqno(4.22) 
$$
where $l_i\in\Zop$ depends on $\nu$. Each of the coefficients of
$e_i$ for $i=1,\ldots,r$ is integer, and since only every $h_0^{th}$
number is divisible by $h_0$, at most ${r\over h_0}+1 $ of them
vanish. Thus it follows that
$$
(\rho - h_0\; \nu)^2 \geq r \left( 1 - {1\over h_0} \right) -1
\,. \eqno(4.23)
$$
Since $h_0\geq 2$, it then follows that the rank of $C_2$ must be
bounded. 

If $\nu$ is a spinor weight, then each $l_i$ is half-odd-integer. Then
the same argument applies, except that $h_0$ is replaced by $h_0/2$ if
$h_0$ is even. For all simply-laced algebras other than $\g_0=$su$(2)$
$h_0\geq 3$, and (4.23) is then still sufficient. In the case of
$\g_0=$su$(2)$, $h_0 l_i$  is then an odd integer, and therefore only
every second coefficient in (4.22) can vanish. This is again
sufficient to conclude that the rank of $C_2$ must be bounded.
\medskip

The above arguments are somewhat abstract, and it may therefore be
instructive to get a better feeling for what the actual bounds
are. For the case where the central node is only linked to the affine
node of $C_1$ and to only one node of each connected component of
$C_2$, we have made a more detailed analysis (that is sketched in
appendix~D). Within this class of constructions, it is shown there
that the rank of an algebra with a principal so(1,2) subalgebra is
always less than $42$. Actually, this bound is probably not attained,
and it would be interesting to find the actual bound. The largest rank
example (within this class of constructions) that we have managed to
construct has rank $19$ and is found by taking $\g=d_{10}$ and
$\g_{2}=e_7$. Its Dynkin diagram is given below.
\ifig\examplepic{The Dynkin diagram of the Kac-Moody algebra $\g$
obtained by taking $\g=d_{10}$ and $\g_2=e_7$.}
{\epsfxsize4.0in\hskip.2cm\epsfbox{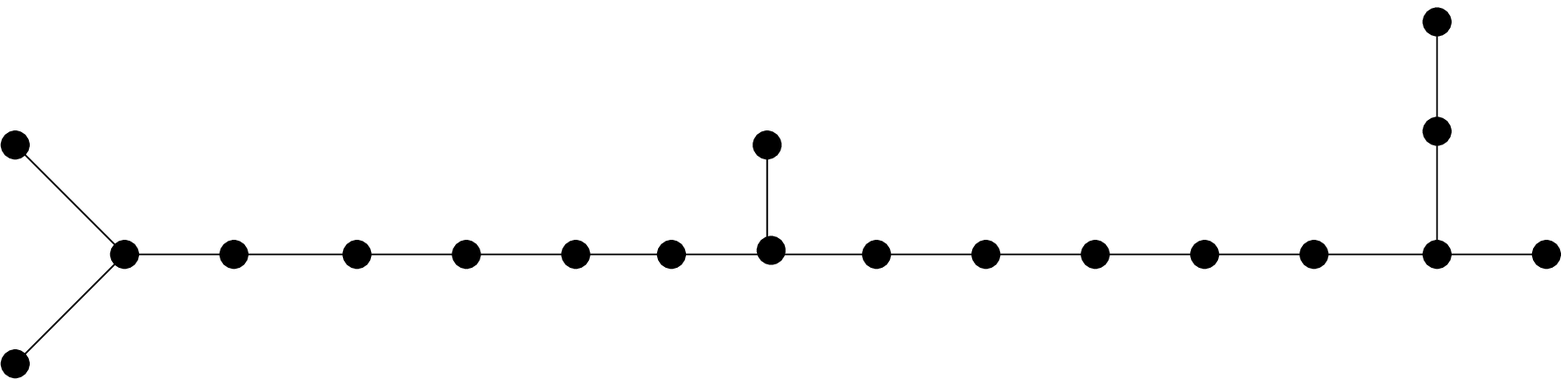}}

\newsec{Very extended Lie algebras}

The Lorentzian Kac-Moody algebras that actually appear in string
theory are rather special examples of the algebras we discussed in
section~4: they arise by joining an affine Kac-Moody algebra
$\g^{(1)}$ via the affine node to a central node that links in turn to
the single finite dimensional Lie algebra $\g_2=$su$(2)$. This special  
construction is a generalisation of the `over-extension' construction
that is explained in \refs{\go}, and we shall therefore call the
resulting Lie algebra `very extended'.  Since these are the examples
of primary interest, it may be worthwhile to describe their structure
in some detail. We shall also {\bf not} assume in the following that
$\g$ has a symmetric Cartan matrix.

Let us begin by considering a finite dimensional semi-simple Lie
algebra $\g$ of rank $r$ whose simple roots $\alpha_i$, 
$i=1,\ldots ,r$ span the lattice $\Lambda_{\g}$. Let us denote the
highest root of $\g$ by $\theta$; we will always normalise the simple
roots of $\g$ such that $\theta^2=2$. This can always be done except
for the Lie algebra $\g=g_2$ which our analysis does not cover. We
choose the convention that the Cartan matrix is defined as 
$A_{ij}={2 (\alpha_i,\alpha_j)\over (\alpha_i,\alpha_i)}$.

\noindent In a first step we enlarge the root lattice of
$\Lambda_{\frak g}$  to be part of
$$
\Lambda_{\frak g} \oplus \Pi^{1,1} \eqno(5.1)
$$
by adding to the simple roots of ${\frak g}$ the extended root
$$
\alpha_0 = k-\theta\,, \eqno(5.2)
$$
where $k \in \Pi^{1,1}\subset \Lambda_{\frak g} \oplus \Pi^{1,1}$ is
described in appendix~B. The corresponding Lie algebra now has $r+1$
simple roots, and it is just the affine Lie algebra of ${\frak g}$
which is often denoted by ${\frak g}^{(1)}$. However, in view of
subsequent developments, we will denote it by ${\frak g}_0$. By
construction we have $(\alpha_0,\alpha_0)=2$ (since $k.k=0$).
Let us denote the scalar products involving the new 
simple root as $(\alpha_0, \alpha_i)\equiv q_i'$ and
$2{(\alpha_i,\alpha_0)\over (\alpha_i,\alpha_i)}\equiv q_i$.  The
corresponding Cartan matrix then has the form
$$
A_{ {\frak g}_0} = \pmatrix{    &      &      &q_1    \cr
   & A_{\frak g}  &      &\vdots   \cr
   &              &      & q_r    \cr
  q_1' & \ldots & q_r' & 2   \cr}\,.
\eqno(5.3)
$$
As has been mentioned before, the determinant of the Cartan matrix
$A_{{\frak g}_0}$ vanishes, $\det\, A_{{\frak g}_0}=0$. 

Clearly, the roots of the affine algebra do not span the whole lattice
$\Lambda_{\frak g} \oplus \Pi^{1,1}$. Rather, the roots of the affine
algebra can be characterised as the vectors $x$ in this lattice which
are orthogonal to $k$, \ie\ $x.k=0$.

We may further extend the affine Lie algebra by adding to the above
simple roots yet another simple root namely \refs{\go}
$$
\alpha_{-1}= -(k+\bar k) \in \ \Lambda_{{\frak g}} \oplus \Pi^{1,1}
\,,
\eqno(5.4)
$$
where we have again used the conventions of appendix~B. We note that
$\alpha_{-1}^2=2$, as well as $(\alpha_{-1},\alpha_0)=-1$ and
$(\alpha_{-1},\alpha_i)=0,\ i=1\ldots ,r$.  The Lie algebra so
obtained is called the over-extended Lie algebra, and we shall denote
it as ${{\frak g}_{-1}}$.  The Cartan matrix associated to the
over-extended Lie algebra has the structure
$$
A_{{\frak g}_{-1}} = \pmatrix{    &      &      &q_1    & 0 \cr
     & A_{\frak g}  &      &\vdots   &    \vdots \cr
     &      &      & q_r & 0 \cr
     q_1' & \ldots &q_r' & 2 & -1 \cr
     0     & \cdots & 0 & -1 & 2 \cr} \,.
\eqno(5.5)
$$
Examining the form of the Cartan matrix we conclude that
$$ \det\, A_{ {\frak g}_{-1}}
= 2 \; \det\, A_{{\frak g}_{0}} - \det\, A_{\frak g}
= - \det\, A_{\frak g}\,.
\eqno(5.6)
$$
Clearly, the root lattice of ${\frak g}_{-1} $ is
$\Lambda _{ {\frak g}_{-1}}=\Lambda_{\frak g}\oplus\Pi^{1,1}$. The
algebra ${\frak g}_{-1}$ is therefore Lorentzian.

\noindent It is possible to enlarge the Lie algebra even further by
considering the lattice
$$
\Lambda_{\frak g} \oplus \Pi^{1,1}\oplus \Pi^{1,1}
= \Lambda_{{\frak g}_{-1}} \oplus \Pi^{1,1}\,.
\eqno(5.7)
$$
We denote the analogue of $k$ and $\bar k$ in the second $\Pi^{1,1}$
lattice by $l$ and $\bar l$, respectively. We now add the new simple
root
$$
\alpha_{-2}= k-(l+\bar l)\,.
\eqno(5.8)
$$
We then have that $(\alpha_{-2},\alpha_{-2})=2$,
$(\alpha_{-2},\alpha_{-1})=-1$, while all other scalar products
involving $\alpha_{-2}$ vanish. Let us denote the resulting Kac-Moody
algebra by ${\frak g}_{-2}$. The corresponding Cartan matrix is then
of the form
$$
A_{ {\frak g}_{-2}} = \pmatrix{    &      &      &q_1    & 0 &0\cr
    & A_{\frak g}  &      &\vdots   &    \vdots &\vdots\cr
    &      &      & q_r & 0 &0\cr
    q_1' & \ldots &q_r' & 2 & -1 &0\cr
    0     & \cdots & 0 & -1 & 2&-1 \cr
    0&\ldots &0&0&-1&2} \,.
\eqno(5.9)
$$
This Cartan matrix is precisely the Cartan matrix that is obtained
from the construction in section~4 with $\g_2= {\rm su}(2)$. Examining
the form of this Cartan matrix we conclude that
$$
\det\, A_{ {\frak g}_{-2}}
= 2\; \det\, A_{ {\frak g}_{-1}} - \det\, A_{{\frak g}_0}
= 2\; \det\, A_{ {\frak g}_{-1}} = - 2\;  \det\, A_{ {\frak g}}\,.
\eqno(5.10)
$$
This is in agreement with (4.6) since the determinant of the Cartan
matrix of su(2) equals $2$ and $\eta$, defined in (4.9), equals
$\eta=1$. The root lattice of ${ {\frak g}_{-2}}$ consists of all
vectors $x$ in $\Lambda_{\frak g}\oplus \Pi^{1,1}\oplus \Pi^{1,1}$
which are orthogonal to the time-like vector
$$
s= l-\bar l= (1,1)\,,
\eqno(5.11)
$$
where we have used the notation of appendix~B. This implies, in
particular, that ${\frak g}_{-2}$ is a Lorentzian algebra.

As before, it is straightforward to calculate the fundamental weights
of the over-extended and very extended algebras. In the over-extended
case the fundamental weights are given as
$$\eqalign{ \lambda_i= & \lambda_i^{\rm f}-(\lambda_i^{\rm f},\theta)\,
(k-\bar k)\,,
\qquad i=1,\ldots ,r\,, \cr
\lambda_0 = & -(k-\bar k)\,, \cr
\lambda_{-1} = & - k\,,} \eqno(5.12)
$$
where $\lambda_i^{\rm f}$ are the fundamental weights of ${\frak g}$. 
On the other hand, the fundamental weights of the very extended
algebra are
$$
\eqalign{
\lambda_i = & \lambda_i^{\rm f}-(\lambda_i^{\rm f},\theta)\,
\left(k-\bar k-{1\over 2}(l+\bar l)\right)\,,
 \qquad i=1,\ldots ,r\,, \cr 
\lambda_0 = & -(k-\bar k-{1\over 2}(l+\bar  l))\,, \cr 
\lambda_{-1} = & -k \,, \cr 
\lambda_{-2} = & - {1\over 2}(l+\bar l) \,.}  
\eqno(5.13)$$ 
It was shown in \refs{\go} that the Weyl vector of an over-extended
algebra is given by 
$$
\rho= \rho_{\rm f} +h\bar{k}-(h+1)k \,,
\eqno(5.14)$$
where $\rho_{\rm f}$ is the Weyl vector of the underlying finite
dimensional Lie algebra $\g$, and $h$ is its Coxeter
number. Similarly, the Weyl vector of the very extended Kac-Moody
algebras is given by 
$$\rho= \rho_{\rm f}+h\bar{k}-(h+1)k-{1\over 2}(1-h) (l+\bar l) \,.
\eqno(5.15)$$

\subsec{Weight lattices}

We now construct the weight lattices of the Kac-Moody algebras
introduced above. For simplicity we shall only consider the
simply-laced case for which the weight lattice $\Lambda_W$ is just the
dual $\Lambda_W=\Lambda_R^\ast$ of the root lattice $\Lambda_R$. These
lattices can be easily found, using the fact that for any  two
lattices $\Lambda_1$ and $\Lambda_2$ we have 
$$
(\Lambda_1\oplus \Lambda_2)^\star
=\Lambda_1^\star\oplus\Lambda_2^\star\,.
\eqno(5.16)
$$
Now the lattice $ \Pi^{1,1}$ is self-dual, and thus the weight lattice
of ${\frak g}_{-1}$ is simply given by
$$
(\Lambda _{{\frak g}_{-1}})^\star= \Lambda _{{\frak g}}^\star
\oplus \Pi^{1,1}\,.
\eqno(5.17)
$$
In particular it follows that
$${(\Lambda _{{\frak g}_{-1}})^\star\over
(\Lambda _{{\frak g}_{-1}})} = \Z_{{\frak g}}\,.
\eqno(5.18)
$$
Given (2.11), this result is consistent with the relation between the
determinants of equation (5.6).
\smallskip

\noindent The weight lattice for ${\frak g}_{-2}$ is given as
$$
(\Lambda _{{\frak g}_{-2}})^\star= \Lambda_{{\frak g}}^\star
\oplus \Pi^{1,1}\oplus \{(r,-r):\ 2r\ \in\ \Zop\}\,,
\eqno(5.19)
$$
where we have used the conventions of appendix~B. In deriving (5.19)
we have noted that the root lattice of ${\frak g}_{-2}$ is
$$
\Lambda_{{\frak g}_{-2}} = \Lambda_{{\frak g}} \oplus \Pi^{1,1}
\oplus \{ (t,-t): t \in \Zop\}  \eqno(5.20)
$$
since $l+\bar{l}=(1,-1)$. The last lattice in (5.19) is generated by
$f=(1/2,-1/2)$ which is not in $\Pi^{1,1}$, but for which
$2f \in \Pi^{1,1}$. Thus we conclude that
$$
{(\Lambda _{ {\frak g}_{-2}})^\star\over
(\Lambda _{ {\frak g}_{-2}})}=\Z_{\frak g} \times \Zop_2\,.
\eqno(5.21)
$$
Here the $\Zop_2$ results from the fact that the dual lattice contains
the vector $f$ in the last factor of $\Pi^{1,1}$. This is consistent
with the factor of $2$ between the two determinants of equation (5.10).

\subsec{The relation to self-dual lattices}

The above extensions were carried out for any finite dimensional
semi-simple Lie algebra ${\frak g}$ of rank $r$, but we now consider
in detail the resulting algebras when $\Lambda_{\frak g}$ is an even
self-dual lattice of dimension $r$, or a sublattice of such a
lattice. Even self-dual Euclidean lattices only exist in dimensions
$D=8n,\ n=1,2,\dots$.

The first non-trivial example of such a lattice occurs in eight
dimensions where there is only one such lattice, the root lattice of
$e_8$. Let us denote the corresponding affine, over-extended and very
extended algebras by $e_9$, $e_{10}$ and $e_{11}$, respectively.
We can choose a basis for the root lattice of $e_8$, $\Lambda_{e_8}$,
to be
$$
\eqalign{
\alpha_1 & = (0,0,0,0,0,1,-1,0)\,, \cr
\alpha_2 & = (0,0,0,0,1,-1,0,0)\,, \cr
\alpha_3 & = (0,0,0,1,-1,0,0,0)\,, \cr
\alpha_4 & = (0,0,1,-1,0,0,0,0)\,, \cr
\alpha_5 & = (0,1,-1,0,0,0,0,0)\,, \cr
\alpha_6 & = (-1,-1,0,0,0,0,0,0)\,, \cr
\alpha_7 & = \left({1\over 2},{1\over 2},{1\over 2},{1\over 2},
{1\over 2},{1\over 2},{1\over 2},{1\over 2}\right)\,, \cr
\alpha_8 & = (1,-1,0,0,0,0,0,0)\,. \cr}
\eqno(5.22)$$
In order to describe the extension and over-extension of $e_8$ we
consider the lattice $\Lambda_{e_8}\oplus \Pi^{1,1}$. The affine root
that gives $e_9$ is now given by
$$
\alpha_0 = k-\theta=\left(-\theta; (1,0)\right) \,,
\eqno(5.23)$$
where $\theta\in\Lambda_{e_8}$ is the highest root of $e_8$, which,
with the above choice of simple roots, is
$$
\theta = (0,0,0,0,0,0,1,-1) \,.
\eqno(5.24)$$
Finally, the over-extended root that enhances this to $e_{10}$ can
then be chosen to be
$$
\alpha_{-1} = -(k+\bar k)=\left( {\bf 0}; (-1,1) \right) \,.
\eqno(5.25)$$
It is easy to see (and in fact well known \refs{\go}) that this
construction gives the root lattice of $e_{10}$.

The lattice $\Lambda_{e_8}\oplus \Pi^{1,1}$ is clearly self-dual by
virtue of equation (5.16). It is of Lorentzian signature and even. Such
lattices only occur in dimensions $D=8n+2,\ n=0,1,2\ldots,$ and the
lattice in each dimension is unique and usually denoted by
$\Pi^{8n+1,1}$. It follows that the root lattice of $e_{10}$ is
precisely this lattice for $n=1$, \ie\ $\Lambda _{e_{10}}=\Pi^{9,1}$.

\noindent Finally, we consider the lattice
$$
\Pi^{9,1} \oplus \Pi^{1,1}
= \Lambda_{e_8} \oplus \Pi^{1,1} \oplus \Pi^{1,1}=\Pi^{10,2} \,,
\eqno(5.26)$$
where the latter lattice is the unique even self-dual lattice of
signature $(10,2)$. The very extended root is given by
$$
\alpha_{-2} = k-(l+\bar l)=\left( {\bf 0} ; (1,0); (-1,1) \right) \,.
\eqno(5.27)$$
The corresponding algebra, $e_{11}$, has been argued to be a
symmetry of M-theory in \refs{\pwtwo}. From equation (5.21) it
now follows that
$${\Lambda^\star_{e_{11}} \over \Lambda_{e_{11}}} = \Zop_2\,.
\eqno(5.28)
$$

\subsec{The $24$-dimensional case}

Next let us consider the extensions of a finite dimensional semi-simple
Lie algebra of rank $24$ whose root lattice is a sublattice of an even
self-dual Euclidean lattice in dimension $24$. In dimension $24$,
there are $24$ such lattices, the so-called Niemeier lattices
\refs{\cs}. One of the Niemeier lattices contains the root lattice
of $d_{24}$, that can be taken to be spanned by the vectors in
$\Zop^{24}$ of the form
$$
\alpha_i = \left(0^{i-1},1,-1,0^{23-i} \right)\,, \qquad 
i=1,\ldots,23\,,
\qquad\qquad
\alpha_{24} = \left(0^{22},1,1\right)\,.
\eqno(5.29)
$$
The root lattice of $d_{24}$ is not self-dual by itself since
$${\Lambda^\star_{d_{24}}\over \Lambda_{d_{24}}}
= \Zop_2 \times \Zop_2\,,
\eqno(5.30)
$$
which is consistent with the fact that $\det\, A_{d_{24}}=4$. The
corresponding self-dual lattice is given by
$$\Lambda^N_{d_{24}}={\Lambda^\star_{d_{24}}\over \Zop_2}\,.
\eqno(5.31)$$
It is obtained from the root lattice of $d_{24}$, $\Lambda_{d_{24}}$,
by adjoining a point of length squared six,
$$
g= \left[ \left({1\over 2}\right)^{24} \right] \,.
\eqno(5.32)
$$
It is easy to see that $g\in \Lambda^\star_{d_{24}}$, and that
$2g\in \Lambda_{d_{24}}$.

Let us denote by $k_{26}$ the over-extension of $d_{24}$ that is
obtained by adding to $d_{24}$ the affine and over-extended roots.
The rank of $k_{26}$ is $26$, and its root lattice is
$$
\Lambda_{k_{26}}=\Lambda_{d_{24}}\oplus \Pi^{1,1}\,.
\eqno(5.33)
$$
It is spanned by the roots of equation (5.29), together with
$$
\eqalign{\alpha_0 & =\left((-1,-1,0^{22});(1,0)\right)\,, \cr
\alpha_{-1} & = \left( (0^{24});(-1,1)\right)\,.}
\eqno(5.34)
$$
We have
$$
{\Lambda_{k_{26}}^\ast \over \Zop_2}=
{\Lambda_{d_{24}}^\ast \over \Zop_2}\oplus \Pi^{1,1}
=\Lambda^N_{d_{24}}\oplus \Pi^{1,1}=\Pi^{25,1} \,,
\eqno(5.35)
$$
since the lattice
${\Lambda_{d_{24}}^\ast\over \Zop_2}\oplus\Pi^{1,1}$ is an
even self-dual lattice of dimensional $26$. It thus follows that
$\Lambda_{k_{26}}$ is contained in $\Pi^{25,1}$.

\noindent Finally we consider the further extension of $d_{24}$ by
considering the lattice
$$\Lambda_{d_{24}}\oplus \Pi^{1,1}\oplus \Pi^{1,1}
\eqno(5.36)
$$
and adding the simple root
$$\alpha_{-2}= \left( (0^{24});(1,0);(-1,1)\right)\,,
\eqno(5.37)
$$
as discussed for the general case above. We shall denote the
corresponding algebra by $k_{27}$; it has been argued to be a
symmetry of the $26$-dimensional closed bosonic string \refs{\pwtwo}. 
It also follows that
$${\Lambda^\star_{k_{27}} \over \Lambda_{k_{27}}}
= \Zop_2 \times \Zop_2 \times \Zop_2 \,.
\eqno(5.38)
$$

\subsec{The $16$-dimensional case}

For completeness, let us conclude this section with a discussion of
the very extended Lie algebra associated to the rank $16$ algebra
$d_{16}$. As in the previous section, the corresponding root lattice
is not self-dual since
$${\Lambda^\star_{d_{16}}\over \Lambda_{d_{16}}}
= \Zop_2 \times \Zop_2\,.
\eqno(5.39)
$$
The corresponding self-dual lattice is given by
$$\Lambda^s_{d_{16}}={\Lambda^\star_{d_{16}}\over \Zop_2}\,.
\eqno(5.40)$$
It is obtained from the root lattice of $d_{16}$, $\Lambda_{d_{16}}$,
by adjoining a point of length squared four,
$$
g= \left[ \left({1\over 2}\right)^{16} \right] \,.
\eqno(5.41)
$$
Let us denote by $m_{18}$ the over-extension of $d_{16}$ that is
obtained by adding to $d_{16}$ the affine and over-extended roots.
The rank of $m_{18}$ is $18$, and its root lattice is
$$
\Lambda_{m_{18}}=\Lambda_{d_{16}}\oplus \Pi^{1,1}\,.
\eqno(5.42)
$$
In particular, we therefore have
$$
{\Lambda_{m_{18}}^\ast \over \Zop_2}=
{\Lambda_{d_{16}}^\ast \over \Zop_2}\oplus \Pi^{1,1}
=\Lambda^s_{d_{16}}\oplus \Pi^{1,1}=\Pi^{17,1} \,,
\eqno(5.43)
$$
and thus $\Lambda_{m_{18}}$ is contained in $\Pi^{17,1}$.

\noindent Finally we consider the further extension of $d_{16}$ by
considering the lattice
$$\Lambda_{d_{16}}\oplus \Pi^{1,1}\oplus \Pi^{1,1}
\eqno(5.44)
$$
and adding the simple root
$$\alpha_{-2}= \left( (0^{16});(1,0);(-1,1)\right)\,,
\eqno(5.45)
$$
as discussed for the general case above. We shall denote the
corresponding algebra by $m_{19}$.

We observe that constructing the very extended Lorentzian Kac-Moody
algebras based on Euclidean self-dual lattices leads to a very
specific set of algebras, namely $e_{11}$, $m_{19}$ and
$k_{27}$. Remakably, $e_{11}$ and $k_{27}$ are thought to be
symmetries of M theory and the $26$-dimensional bosonic string
\refs{\pwtwo}. The above construction also makes it clear that the
symmetries of these two theories are related to the unique even
self-dual Lorentzian lattices in ten and $26$ dimensions,
respectively. These observations encourage the speculation that there
should also exist a $18$-dimensional string with a symmetry
$k_{19}$. These and other implications for string theory will be
discussed elsewhere.

\subsec{Principal so(1,2) subalgebras}

It may also be interesting to analyse which of the over-extended and
very extended Kac Moody algebras possess a principal so(1,2)
subalgebra. As before we have to analyse the condition of equation
(A.8). Using (A.9), as well as the explicit expressions for the
fundamental weights given in (5.12) and (5.13), we find that for an
over-extended Kac-Moody algebra the left hand side of equation (A.8)
is 
$$\eqalign{
\sum_a A_{a-1}^{-1} & =-h\,, \cr
\sum_a A_{a0}^{-1}  & =-(2h+1)\,, \cr
\sum_a A_{aj}^{-1}  & =-{1\over 2}\, n_j\, (\alpha_j,\alpha_j)\,
(2h+1) +\sum_i A_{{\rm f}\ ij}^{-1} \,.}
\eqno(5.46)$$
Here $h$, $n_i$, $A_{{\rm f}\ ij}^{-1}$ are the Coxeter number, the
Kac labels, and the inverse Cartan matrix of the finite
dimensional Lie algebra ${\frak g}$, respectively.

\noindent Similarly, we find for the case of the very extended
Kac-Moody algebra
$$\eqalign{
\sum_a A_{a-2}^{-1} & = -{1\over 2}(h-1)\,, \cr
\sum_a A_{a-1}^{-1} & = - h\,, \cr
\sum_a A_{a0}^{-1}  & = - {3\over 2}(h+1)\,, \cr
\sum_a A_{aj}^{-1} & = - {3\over 4}\, n_j\, (\alpha_j,\alpha_j)\,
(h+1) +\sum_i A_{{\rm f}\ ij}^{-1} \,.}
\eqno(5.47)$$
\par
We observe that in both cases only the sums $\sum_a A_{aj}^{-1}$
do not automatically satisfy the required condition of equation
(A.8).  The relevant condition depends therefore on the Kac
labels, Coexter numbers, and the corresponding sums in the finite
dimensional Lie algebra. For the case of the classical Lie algebras,
the relevant data have been collected in appendix~C.
\smallskip

As an example, let us consider the case of the very extended $a_n$
algebra in more detail. Using equation (C.9) of appendix~C we find
that
$$
\sum_a A_{aj}^{-1}
=-{3\over 2}\, (n+2)+{j\over 2}\, (n+1-j) \,.
\eqno(5.48)$$
This has its maximum  when $j={n+1\over 2}$ for $n$ odd and
$j={n\over 2}$ for $n$ even. In the first case the maximum of
$\sum_a A_{aj}^{-1}$ is ${1\over 8}(n^2-10n-23)$
while in the latter case it is ${1\over 8}(n^2-10n-24)$. Hence
$\sum_a A_{aj}^{-1}$ is non-positive if and only if $n\le 12$, and
thus a principal so(1,2) subalgebra exists for the very extended $a_n$ 
algebra if $n\le 12$. We  summarise the results for all over-
and very extended Lie algebras in the following table.
\vskip 0.4cm
\vbox{
\centerline{\vbox{
\hbox{\vbox{\offinterlineskip
\def\tablespace{height2pt&\omit&&\omit&& \omit&\cr}

\def\tableruleA{\tablespace\noalign{\hrule height1pt}\tablespace}
\hrule\halign{&\vrule#&\strut\hskip0.2cm\hfil#\hfill\hskip0.2cm\cr
\tablespace
& && \hbox{over-extended} && {very extended}  &\cr
\tableruleA
& $a_n$ && $n\leq 16$ && $n\leq 12$ &\cr \tablespace
& $d_n$ && $n\leq 16$ && $n\leq 12$ &\cr \tablespace
& $e_n$ && $n=6,7,8$ && $n=6,7,8$ &\cr \tablespace
& $b_n$ && $n\leq 15$ && $n\leq 11$ &\cr \tablespace
& $c_n$ && $n\leq 8$ && $n\leq 6$ &\cr \tablespace
& $f_4$ && yes && yes  &\cr \tablespace
& $g_2$ && yes && yes  &\cr
\tablespace}\hrule}}}}
\centerline{
\hbox{{\bf Table 1:} {\it The algebras with principal so(1,2)
subalgebras.}}}}

The algebras of particular interest to string theory are $e_{11}$ and
$k_{27}$. These algebras are the very extended algebras corresponding
to  $e_8$ and $d_{24}$, respectively. The above table implies that
while $e_{11}$ admits a principal so(1,2) subalgebra, $k_{27}$ does
not. We also note that the other algebra that is related to self-dual
lattices, $m_{19}$ (see section~5.4), also does not admit a principal
so(1,2) subalgebra since it is the very extended algebra corresponding
to $d_{16}$.

\newsec{Other constructions}

Up to now we have discussed Lorentzian Kac-Moody algebras that arise
by means of a certain simple construction. While these Kac-Moody
algebras may be preferred  in some way, it is clear that they
do not account for {\it all} Lorentzian Kac-Moody algebras, and not
even for all those with a principal so(1,2) subalgebra.

In this section we want to describe some other classes of Lorentzian
Kac-Moody algebras that can be obtained by similar types of
constructions. In each case we shall also analyse for which examples
the resulting algebra has a principal so(1,2) subalgebra.

\subsec{Adding a different node}

The simplest modification of the above construction leading to a very 
extended Lie algebra is to attach the very extended node at a
different place in the Dynkin diagram of the over-extended algebra. As
before, we shall take the roots to belong to the lattice
$$
\Lambda_{\frak g} \oplus \Pi^{1,1}\oplus \Pi^{1,1}
= \Lambda_{{\frak g}_{-1}} \oplus \Pi^{1,1}\,.
\eqno(6.1)
$$
We take the roots of our new algebra to be the roots of the
over-extended algebra (see section~5), except that we replace
$\alpha_{i_0}$ by $\widehat\alpha_{i_0}=\alpha_{i_0}+l$, where $l$ is
defined as in section~5 and $i_0$ is a chosen index on the Dynkin
diagram; in addition we choose
$\alpha_{-2}=-(l+\bar l)$. The corresponding Dynkin diagram is then the
diagram that is obtained from the Dynkin diagram of the over-extended
algebra by adding a node that is attached to the $i_0^{th}$ node.

\noindent The roots of this new algebra are orthogonal to the vector
$$s=(l-\bar l)+2{\lambda_{i_0}\over (\alpha_{i_0},\alpha_{i_0})}
+2{(\lambda_{i_0},\theta)\over (\alpha_{i_0},\alpha_{i_0})}\, 
(\bar k-k) \,.
\eqno(6.2)$$
The resulting algebra is therefore  Lorentzian if $s$ is time-like,
\ie\ if
$$ 4{(\lambda_{i_0},\lambda_{i_0})\over
(\alpha_{i_0},\alpha_{i_0})^2} < 2\, \left (1+ n_{i_0}^2 \right) \,.
\eqno(6.3)$$
For example, if we take ${\frak g}=a_n$,  then the algebra is
Lorentzian if
$$ i_0 (n+1- i_0)< 4(n+1) \,.
\eqno(6.4)$$
We have also analysed (using {\tt Maple}) which of these algebras have
a principal so(1,2) subalgebra. We have found that for $n=16$, two of
the algebras so obtained have a principal so(1,2) subalgebra; their
Dynkin diagrams are shown below.

\ifig\exampletwopic{Dynkin diagrams of Lorentzian Kac-Moody algebras
with a principal so(1,2) subalgebra that cannot be obtained by any of
the constructions described in section~3 and 4.}
{\epsfxsize1.5in\epsfbox{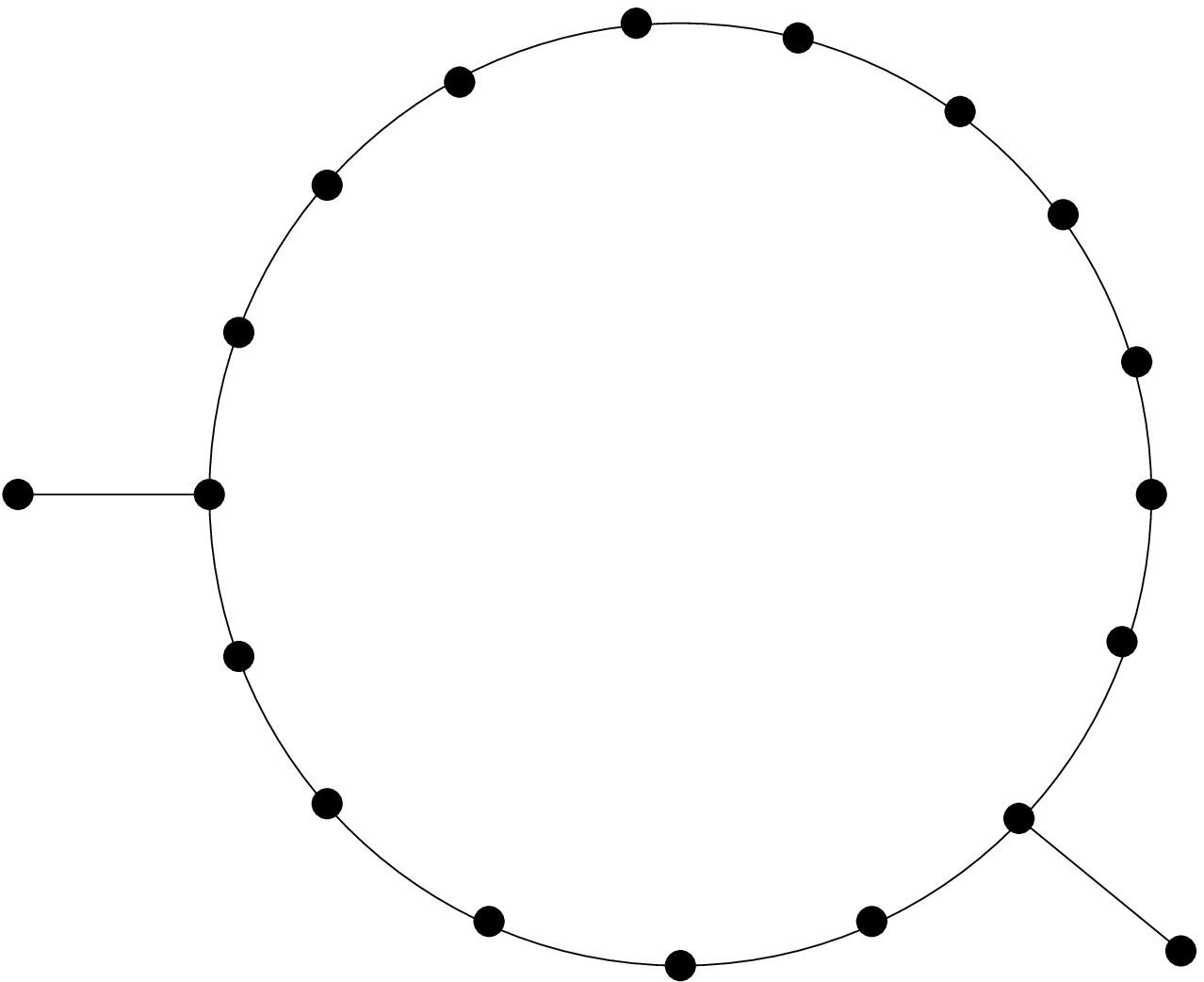}\hskip4cm
\epsfxsize1.4in\epsfbox{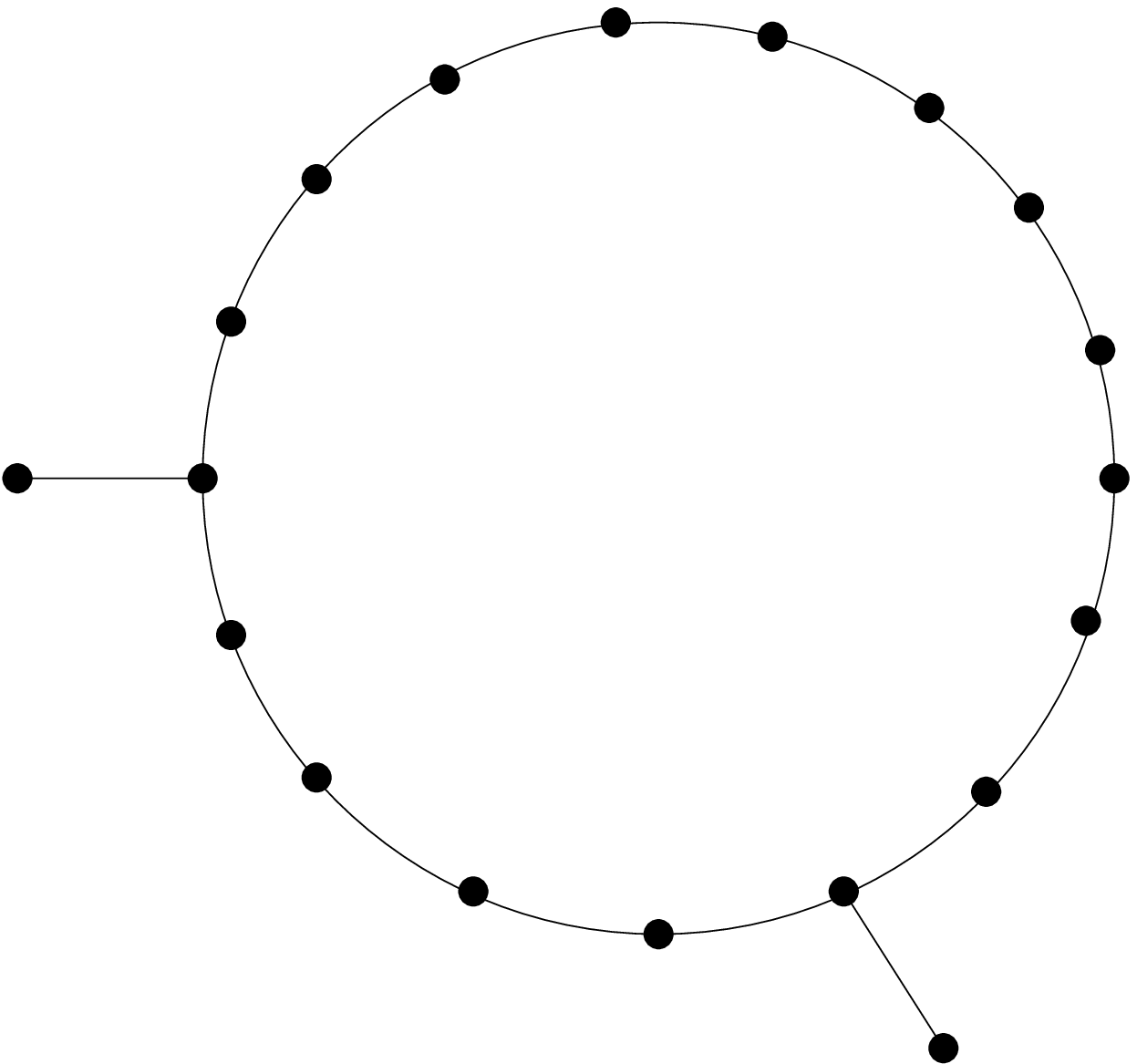}}

\subsec{Symmetric fusion}

The construction of section~4 is somewhat asymmetric in that one
affine Kac-Moody algebra is singled out. In this section we want to
describe a more symmetrical construction that also gives rise to a
Lorentzian Kac-Moody algebra. As will become apparent, this
construction can actually be regarded as a special case of the
construction of section~4. However, it may nevertheless be interesting
to discuss it in its own right.

Suppose ${\frak g}_1$ and ${\frak g}_2$ are two finite dimensional
simply-laced simple Lie algebras of rank $r_1$ and $r_2$,
respectively. We want to construct a Lorentzian algebra
${\frak g}_1\diamond {\frak g}_2$ whose rank is $r_1+r_2+2$. The root
lattice of this algebra will be given by
$$\Lambda _{{\frak g}_1}\oplus \Pi^{1,1}\oplus \Lambda_{{\frak g}_2}\,.
\eqno(6.5)$$
We take the simple roots of the algebra
${\frak g}_1 \diamond {\frak g}_2$ to be those of ${\frak g}_1$ and
${\frak g}_2$, which we denote by $\alpha_i$, $i=1,\ldots, r_1$ and
$\beta_{j}$, $j=1,\ldots, r_2$, respectively. We add to these two
further simple roots,
$$\alpha_0=k-\theta_1 \qquad {\rm and } \quad
\beta_{0}=-\bar k-\theta_2  \,,
\eqno(6.6)$$
where $k$ and $\bar k$ belong to $\Pi^{1,1}$, as explained in
appendix~B, and $\theta_1$ and $\theta_2$ are the highest roots of
${\frak g}_1$ and ${\frak g}_2$, respectively. Since
${\frak g}_1$ and ${\frak g}_2$ are simply-laced algebras, we have
$\alpha_0^2=\beta_0^2=2$, $(\alpha_0,\beta_0)=-1$. The corresponding
Cartan matrix is then given by
$$
A_{ {\frak g}_1\diamond {\frak g}_2} =
\pmatrix{    &      &      & q_1 & 0&\ldots
&\ldots&0\cr & A_{{\frak g}_1}  &      &\vdots   &    \vdots
&\vdots&\vdots&\vdots \cr
   &      &      & q_{r_1} & 0&\ldots&\ldots &0\cr
    q_1 & \ldots &q_{r_1} & 2 & -1 &0&\dots&0\cr
    0     & \cdots & 0 & -1 & 2&p_1&\ldots &p_{r_2}\cr
    \vdots & & \vdots &   0&p_1  &  \cr
    \vdots & &\vdots &\vdots&\vdots & &A_{{\frak g}_2}\cr
    0&\ldots &0& 0 &p_{r_2}& & }\,,
\eqno(6.7)
$$
where $q_i=(\alpha_i,\alpha_0)$,
and $p_j=(\beta_j,\beta_0)$. By construction,
it is clear that the Dynkin diagram of 
${\frak g}_1 \diamond {\frak g}_2$ contains the Dynkin diagrams of
${\frak g}^{(1)}_1$ and ${\frak g}^{(1)}_2$, respectively. In fact, it
is obtained by joining the two affine diagrams with a single line
between the two affine roots. If we think of the affine node of either 
${\frak g}^{(1)}_1$ or ${\frak g}^{(1)}_2$ as the central node, 
the Dynkin diagram is then of the form described in section~4.

By considering the column associated with the root $\alpha_0$ we can
calculate the determinant of the corresponding Cartan matrix, and we
find as before that
$$\det\, A_{{\frak g}_1 \diamond {\frak g}_2}=
-\det \, A_{{\frak g}_1} \; \det\, A_{{\frak g}_2} \,.
\eqno(6.8)$$
This is in agreement with (4.6). It also follows from the analysis of
section~4 that the Kac-Moody algebra
${\frak g}_1 \diamond {\frak g}_2$ is Lorentzian.

Next we want to analyse for which cases this Lorentzian algebra has a
principal so(1,2) subalgebra. As before, we can determine the weights,
and we find that they are given by
$$\eqalign{
\ell_i^{(1)} & =\lambda_i^{(1)}+(\lambda_i^{(1)},\theta_1)\, \bar k\,, 
\qquad   \ell_{\alpha_0}=\bar k\,, \cr
\ell_i^{(2)} & =\lambda_i^{(2)}-(\lambda_i^{(2)},\theta_2)\, k\,, 
\qquad   \ell_{\beta_0}=- k\,,}
\eqno(6.9)$$
where $\lambda_i^{(p)}$ are the fundamental weights of $\g_p$,
$p=1,2$. \noindent It is straightforward to show that
$$\sum_a 
\left(A_{{\frak g}_1 \diamond {\frak g}_2}\right)^{-1}_{a\ \alpha_0}
=-h(\g_2) < 0\,, \qquad
\sum_a \left(A_{{\frak g}_1 \diamond {\frak g}_2}\right)^{-1}_{a j}
=- n_j^{(1)} \,h(\g_2) +  \sum_{i=1}^{r_1}
A^{-1}_{\g_1\, ij} \,,
\eqno(6.10)$$
where $j$ denotes a node of $\g_1$. The sums over the other columns
may be obtained from the above by exchanging $1\leftrightarrow 2$. As
in our discussion of section~5.5 we therefore conclude that a
principal so(1,2) subalgebra exists if the second sum in equation
(6.10) is also negative. Let us discuss a few examples in detail.

For the case of ${\frak g}_1=a_{n_1}$, ${\frak g}_2=a_{n_2}$ we find,
using the results of appendix~C, that there is a principal so(1,2)
subalgebra if $(n_1+1)^2\le 8 (n_2+1)$ and $(n_2+1)^2\le 8(n_1+1)$. (In
deriving these inequalities we have assumed that both $n_1$ and $n_2$
are odd, but similar inequalities also hold if $n_1$ or $n_2$ are even.)
By squaring the first condition and using the second we conclude that  
$n_1\le 7,\ n_2\le 7$. It is straightforward to find the solutions
(taking, without loss of generality, $n_2\geq n_1$): apart from the
case $n_1=n_2= 1,\ldots, 7$ and $n_2=n_1+1=1,\ldots ,5$
there is the one additional solution $n_2=3, n_1=1$.

Similarly we find that for the case of ${\frak g}_1=d_{n_1}$, 
${\frak g}_2=d_{n_2}$ there exists a principal so(1,2) subalgebra if 
$n_1=n_2= 3,\ldots , 8$,  $n_2=n_1+1$, $n_1=3,\ldots ,7$
and $n_2=n_1+2=4,5,6$. Also, for the case of 
${\frak g}_1=a_{n_1}$, ${\frak g}_2=d_{n_2}$ we find 
that there exists a principal so(1,2) subalgebra only if
$n_1\le 7$ and $n_2\le 6$.

Finally, we have checked that the algebras $a_n\diamond e_m$ only
have a principal so(1,2) subalgebra provided that $m=6$ and $n=7,8$,
and that the algebras $d_n\diamond e_m$ only have a principal so(1,2)
subalgebra provided that $m=6$, $n=5,6,7$ or $m=7$,
$n=8,9$. Furthermore the algebras  $e_n\diamond e_m$ have a principal
so(1,2) subalgebra if and only if $n=m$ with $n=6,7,8$.
\bigskip

It follows from (6.5) and (6.8) that the symmetric fusion of two
finite dimensional simple Lie algebras gives rise to an even self-dual
root lattices provided that the two finite dimensional
Lie algebras are separately self-dual. The only example is therefore
the rank $18$ Lorentzian algebra $e_8\diamond e_8$. Another
interesting example is the algebra $e_8\diamond d_{16}$, that actually
equals the algebra $k_{26}$ discussed earlier. The root lattice of
this algebra is not self-dual, but as explained in section~5, can be
made self-dual by the addition of a spinor weight of $d_{16}$.

\newsec{Conclusions}

In this paper we have described and analysed a certain subclass of
(Lorentzian) Kac-Moody algebras that are in many respects rather
amenable to a general analysis. These algebras are characterised by
the property that their Dynkin diagram contains at least one node,
upon whose  deletion the diagram becomes that of a direct sum of
affine and finite Lie algebras. We have described the conditions under
which these algebras are actually Lorentzian, and we have given
explicit descriptions for their simple roots and fundamental
weights. We have also found simple formulae for the determinants of
the corresponding Cartan matrices. Using similar techniques one can
derive their characteristic polynomials, thus reproducing (for
the case of the finite Lie algebras) known results in a rather 
elegant fashion.

We have discussed the Lorentzian algebras whose root lattices are
self-dual. In particular, we have shown how to construct, for a given
even self-dual Lorentzian lattice, a large number of inequivalent 
algebras whose root lattice is the given Lorentzian lattice. Finally 
we have studied the question of whether our Lie algebras possess a
principal so(1,2) subalgebra.   

A special subclass of the algebras we have considered are what we
called very extended Lie algebras. These very extended algebras arise
as symmetries of M-theory and the bosonic string \refs{\pwtwo}, thus
suggesting that the subclass of algebras described in this paper may
play an important r\^ole in physics.

The methods we have described in this paper will probably generalise
to other classes of algebras. In particular, one can use for example 
our determinant formulae iteratively to analyse Dynkin diagrams that 
reduce to that of affine and finite Lie algebras upon deletion of two
or mode nodes, {\it etc.} It would be interesting to explore these
ideas further.

\vskip 1cm

\centerline{{\bf Acknowledgements}}\pano

We thank Hermann Nicolai and Andrew Pressley for useful
conversations. MRG is supported by the Royal Society via a University 
Research Fellowship. He is grateful to the Isaac Newton Institute
for hospitality while this paper was being completed. DIO and PCW are
grateful to CERN for hospitality during an early stage of this work. 
MRG, DIO and PCW acknowledge partial support from 
the EU network `Superstrings' (HPRN-CT-2000-00122), and MRG and PCW
acknowledge partial support from the PPARC special grant `String
Theory and Realistic Field Theory', PPA/G/S/1998/0061.

\appendix{A}{Principal so(1,2) subalgebras}

It is well known that every finite dimensional semi-simple Lie algebra 
contains a principal so(3) subalgebra. In reference \refs{\no} it was
shown that this result generalises to hyperbolic Kac-Moody algebras in
the sense that they contain a real principal so(1,2)
subalgebra. However, as we shall see, this property is not just
restricted to hyperbolic algebras, but holds for a wider class of
Lorentzian algebras, including many of the algebras constructed in
this paper. 

Let us recall the discussion of reference \refs{\no}, extended to
include non-simply-laced Lie algebras. We define a generator $J_3$ by
$J_3=-\rho^iH_i$, where $H_i$ are the Cartan generators in the
Cartan-Weyl basis, and $\rho$ is the Weyl vector which, by definition,
satisfies $(\rho,\alpha_i)=1$ for all simple roots $\alpha_i$. If we
were dealing with a finite dimensional semi-simple Lie algebra, the
definition of $J^3$ would not include a minus sign and we would find
an so(3) subalgebra.

Next we recall that the fundamental weights $\lambda_j$ of a Kac-Moody
algebra are characterised by the property
$$
{2 (\lambda_j,\alpha_i)\over (\alpha_i,\alpha_i)}=\delta_{ij} \,.
\eqno(\hbox{A.1})$$
Given this definition, we can therefore express the Weyl vector as
$$\rho=\sum_{i=1}^r {2\over (\alpha_i,\alpha_i)}\, \lambda_i\,.
\eqno(\hbox{A.2})$$
Furthermore, using the definition of the Cartan matrix
$A_{ij}=2{(\alpha_i,\alpha_j)\over (\alpha_i,\alpha_i)}$ we find that
$$\lambda_i=\sum_{j=1}^r (\alpha_i,\alpha_i) A^{-1}_{ij}
{\alpha_j\over (\alpha_j,\alpha_j)}\,,
\eqno(\hbox{A.3})$$
and, as a result, we can express the Weyl vector as
$$\rho=\sum_{i,j} A^{-1}_{ij} {2\alpha_j\over
(\alpha_j,\alpha_j)}\,.
\eqno(\hbox{A.4})$$
The generator $J_3$ then becomes
$$J_3=-\sum_{i,j} A^{-1}_{ij}\, H_j\,,
\eqno(\hbox{A.5})$$
where now the $H_j$ are the Cartan generators in the Chevalley basis.
\par
The remaining generators of the so(1,2) algebra, denoted  $J^\pm$,
are taken to be given by sums of the simple positive and negative root
generators respectively,
$$J^+=\sum_{i=1}^r p_i E_i\,,\qquad J^-=\sum_{i=1}^r q_i F_i\,,
\eqno(\hbox{A.6})$$
where $p_i, q_i$ will be determined shortly. Given the hermiticity
property $(E_i)^\dagger=F_i$, the generators $J^+$ and $J^-$ inherit
the standard hermiticity property $(J^+)^\dagger=J^-$ provided that 
$p_i=q_i^\ast$. On the other hand, demanding that $[J^+,J^-]=-J^3$,
and using the Serre relations, one finds that 
$$p_j\, q_j = |q_j|^2 = -\sum_{i=1}^r A_{ij}^{-1} \,.
\eqno(\hbox{A.7})$$
The remaining relations of so(1,2) $[J^3,J^\pm]=\pm J^\pm$
are then automatically satisfied. Hence we conclude that a real
principal so(1,2) subalgebra exists if and only if
$$\sum_{i=1}^r A_{ij}^{-1}\leq 0\qquad \hbox{for each $j$}\,.
\eqno(\hbox{A.8})$$
We also note that it follows from equation (A.3) that 
$$
A_{ij}^{-1}={2\over (\alpha_i,\alpha_i)}(\lambda_i,\lambda_j)\,.
\eqno(\hbox{A.9})$$
In the following we shall only consider principal so(1,2) subalgebras
that satisfy the above reality property; we shall therefore drop the 
qualifier `real'.

\appendix{B}{Some properties of $\hbox{II}^{D-1,1}$}

Even self-dual Lorentzian lattices only exist in dimensions
$D=8n+2, n=0,1,2,\dots,$ and for each $n$, there is only one such
lattice, denoted by $\II^{D-1,1}$. These lattices can be defined as
follows. Let $x=(x^1,\ldots ,x^{D-1};x^0)\in\Rop^{D-1,1}$, and
let us denote by $r\in\Rop^{D-1,1}$ the vector
$r=({1\over 2},\ldots ,{1\over 2};{1\over 2})$. Then $x\in\II^{D-1,1}$
provided that
$$x.r\ \in\ \Zop\eqno(\hbox{B.1})$$
and, in addition, either
$$ {\rm all} \ x^\mu \in \Zop\qquad  {\rm or\qquad  all} \
x^\mu- r^\mu  \in \Zop\,.
\eqno(\hbox{B.2})
$$
In the above, the scalar product is the usual scalar product of
Minkowski space, \ie\
$$
x.y= \sum_{i=1}^{D-1} x^iy^i-x^{0}y^{0}\,. \eqno(\hbox{B.3})
$$

Let us now consider in detail the lattice $\II^{1,1}$. Using equations
(B.1) and (B.2) it is straightforward to verify that $\II^{1,1}$
consists of the vectors
$$ (m,2p+m) \quad {\rm and }\quad
\left(n+{1\over 2}, 2q+n+{1\over 2}\right) \quad
\forall\ m,n,p,q\ \in\ \Zop \,. \eqno(\hbox{B.4})
$$
In this paper we will use a description of the lattice $\II^{1,1}$
in terms of vectors $z=(z^+, z^-)$ that are related to the vectors $x$
given above by the change of basis
$$ z^+= x^0+x^1\,,\qquad
  z^-= {1\over 2}(x^0-x^1)\,.
\eqno(\hbox{B.5})
$$
In terms of these vectors the scalar product becomes
$x.y=-z^+w^- - z^-w^+$, where $w^\pm$ are defined in terms of $y$ as in
(B.5). In the basis described by $(z^+,z^-)$, the vectors
of the lattice $\II^{1,1}$ have the simple form
$$(n,m)\quad \forall\ n,m\ \in \Zop\,.
\eqno(\hbox{B.6})$$
The vector $r$ is now simply $r=(1,0)$.

The null vectors of $\II^{1,1}$ are clearly of the form $(n,0)$ and
$(0,m)$ and so the  primitive null vectors can be taken to be
given by $k\equiv(1,0)$ and $\bar k\equiv(0,-1)$. We have chosen
these vectors such that $k.\bar k=1$. Clearly, all vectors of the
lattice $\II^{1,1}$ are of the form
$pk+q \bar k$ where $p,q \in \Zop$. There are only two points of
length squared two in $\II^{1,1}$,  namely $\pm (k+\bar k)$.

\appendix{C}{Roots and weights of the classical Lie algebras}

In this appendix we list the roots, weights, inverse Cartan matrices
and some other properties of the classical finite dimensional
simple Lie algebras. 
Recall that a finite dimensional simple Lie algebra $\g$ possesses a 
highest root $\theta$ which can be written in terms of the simple
roots $\alpha_i$ as
$$\theta=\sum_{i=1}^r n_i \, \alpha_i\,.
\eqno(\hbox{C.1})$$
We refer to the $n_i$ as the Kac labels. The Coxeter number $h(\g)$ is 
related to the height of the highest root by
$$h(\g)=1+\sum_{i=1}^r n_i \,.
\eqno(\hbox{C.2})$$
The Kac labels can also be expressed as
$$
n_i= { 2 (\theta, \lambda_i)\over (\alpha_i,\alpha_i)} \,.
\eqno(\hbox{C.3})$$
In our conventions the Cartan matrix is given by
$A_{ij}=2{(\alpha_i,\alpha_j)\over (\alpha_i,\alpha_i)}$,
and the fundamental weights are defined by
$${2 (\lambda_j,\alpha_i)\over (\alpha_i,\alpha_i)}=\delta_{ij} \,.
\eqno(\hbox{C.4})$$
The inverse Cartan matrix can be expressed in terms of these by
$$A_{ij}^{-1}={2 \over (\alpha_i,\alpha_i) }\, (\lambda_i,\lambda_j)
\,. \eqno(\hbox{C.5})$$

\subsec{The algebras $a_r$ or $su(r+1)$}

\noindent Let $e_i$, $i=1,\ldots, r+1$, be a set of pairwise
orthogonal unit vectors in $\Rop^{r+1}$. We can write the roots of
$su(r+1)$ as
$$\alpha_i=e_i-e_{i+1} \,, \qquad i=1,\ldots, r\,.
\eqno(\hbox{C.6})$$
The highest root is $\theta=e_1-e_{r+1}=\sum_{i=1}^{r}\alpha_i$.
Hence, the Kac labels are given by $n_i=1$ and the Coxeter number
is $h=r+1$. The fundamental weights are given by
$$\lambda_i=\sum_{j=1}^{i} e_j - {i\over r+1} \sum_{j=1}^{r+1} e_j \,.
\eqno(\hbox{C.7})$$
In particular, we therefore have that the Weyl vector is given by
$$
\rho = \sum_{i=1}^{r+1} \left( {r+2-2i \over 2}\right) e_i \,.
\eqno(\hbox{C.8})
$$
Using equation (C.5) we find that the inverse Cartan
matrix is given by
$$A_{ij}^{-1}=i{(r+1-j)\over r+1}\,, \qquad \hbox{where  $j\ge i$.}
\eqno(\hbox{C.9})$$
Summing on the first index we find that
$$\sum_{i=1}^r A_{ij}^{-1}= {j\over 2}(r+1-j) \,.
\eqno(\hbox{C.10})$$

\subsec{The algebras $d_r$ or $so(2r)$}

\noindent Let $e_i$, $i=1,\ldots, r$, be a set of orthonormal vectors
in $\Rop^r$. The roots of $so(2r)$ are given by
$$\eqalign{
\alpha_i & =e_i-e_{i+1}\,, \qquad i=1,\ldots ,r-1\,, \cr
\alpha_r & =e_{r-1}+e_r\,.}
\eqno(\hbox{C.11})$$
The highest root is
$$\theta=e_1+e_2=\alpha_1+2\sum_{i=2}^{r-2}\alpha_i+\alpha_{r-1}
+\alpha_r \,.
\eqno(\hbox{C.12})$$
Hence, the Kac labels are given by $n_i=2$, $i=2,\ldots,r-2$,
$n_1=n_{r-1}=n_r=1$, and the Coxeter number is $h=2(r-1)$. The
fundamental weights are given by
$$\eqalign{
\lambda_i & =\sum_{j=1}^{i} e_j\,, \qquad i=1,\ldots ,r-2\,, \cr
\lambda_{r-1} & ={1\over 2}\sum_{j=1}^{r-1} e_j- {1\over 2} e_r\,, \cr
\lambda_r & ={1\over 2}\sum_{j=1}^{r-1} e_j+{1\over 2} e_r\,.}
\eqno(\hbox{C.13})$$
The Weyl vector is therefore of the form
$$
\rho = \sum_{i=1}^{r-1} (r-i)\, e_i \,.
\eqno(\hbox{C.14})
$$
Using equation (C.5) we find that the inverse Cartan
matrix is then
$$\eqalign{ A_{ij}^{-1} & = i\,,\qquad \qquad\quad\;\;
\hbox{for $i\leq j,\; i,j=1,\ldots ,r-2\,,$} \cr
A_{i\ r-1}^{-1} & = A_{i\ r}^{-1} = {i\over 2}\,, \qquad
\hbox{for $i=1,\ldots,r-2\,,$}\cr
A_{r-1\ r-1}^{-1} & =A_{r\ r}^{-1} ={r\over 4}\,, \cr
A_{r-1\ r}^{-1} & = {r-2\over 4} \,.}
\eqno(\hbox{C.15})$$
Summing on the first index we find that
$$\sum_{i=1}^r A_{ij}^{-1}= {j\over 2}(2r-1-j)\,, \qquad
j=1,\ldots ,r-2\,,
\eqno(\hbox{C.16})$$
and
$$\sum_{i=1}^r A_{i\ r-1}^{-1}= \sum_{i=1}^r A_{ir}^{-1}=
{r\, (r-1)\over 4}\,.
\eqno(\hbox{C.17})$$

\subsec{The algebras $b_r$ or $so(2r+1)$}

\noindent Let $e_i$, $i=1,\ldots, r$, be a set of orthonormal vectors
in $\Rop^{r}$. The roots of $so(2r+1)$ are given by
$$\eqalign{
\alpha_i & = e_i-e_{i+1}\,, \qquad i=1,\ldots ,r-1\,, \cr
\alpha_r & =e_r\,.}
\eqno(\hbox{C.18})$$
The highest root is
$$\theta=e_1+e_2=\alpha_1+2\sum_{i=2}^{r}\alpha_i\,.
\eqno(\hbox{C.19})$$
Hence, the Kac labels are given by $n_i=2$, $i=2,\ldots,r$ and
$n_1=1$, and  the Coxeter number is $h=2r$. The fundamental weights
are 
$$\eqalign{
\lambda_i & =\sum_{j=1}^{i} e_j\,, \qquad i=1,\ldots ,r-1\,, \cr
\lambda_r & = {1\over 2} \sum_{j=1}^{r} e_j \,.}
\eqno(\hbox{C.20})$$
Using equation (C.5) we find that the inverse Cartan
matrix is given by
$$\eqalign{
A_{ij}^{-1} & =i\,, \qquad \hbox{for $i\leq j$, $i,j=1,\ldots ,r-1\,,$} 
\cr 
A_{r\ i}^{-1} & = i\,, \cr
A_{i\ r}^{-1}& ={i\over 2}\,, \cr
A_{r\ r}^{-1}& ={r\over 2}\,.}
\eqno(\hbox{C.21})$$
Summing on the first index we find that
$$\eqalign{
\sum_{i=1}^r A_{ij}^{-1} & = {j\over 2}(2r+1-j)\,, \qquad
j=1,\ldots ,r-1\,, \cr
\sum_{i=1}^r A_{ir}^{-1} & = {r\over 4}(r+1)\,.}
\eqno(\hbox{C.22})$$

\subsec{The algebras $c_r$ or $sp(2r)$}

\noindent Let $e_i$, $i=1,\ldots, r$ be a set of orthonormal vectors
in $\Rop^{r}$. The roots of $sp(2r)$ are
$$\eqalign{
\alpha_i & ={1\over \sqrt 2}(e_i-e_{i+1})\,, \qquad i=1,\ldots ,r-1\,,
\cr
\alpha_r & =\sqrt{2} e_r\,.}
\eqno(\hbox{C.23})$$
The highest root is
$$\theta=\sqrt 2 e_1=2\sum_{i=1}^{r-1}\alpha_i+\alpha_r \,.
\eqno(\hbox{C.24})$$
Hence, the Kac labels are given by $n_i=2$, $i=1,\ldots, r-1$ and
$n_r=1$, and the Coxeter number is $h=2r$. The fundamental weights are
$$\lambda_i={1\over \sqrt 2}\sum_{j=1}^{i} e_j\,, \qquad
i=1,\ldots ,r \,.
\eqno(\hbox{C.25})$$
Using equation (C.5) we find that the inverse Cartan matrix is then
$$\eqalign{
A_{ij}^{-1} & =i\,,\qquad \hbox{for $j\ge i$, $i,j=1,\ldots ,r-1$}\,,
      \cr 
A_{i r  }^{-1} & = i\,,\cr
A_{r i}^{-1}& ={i\over 2}\,,\cr
A_{rr}^{-1}& ={r\over 2}\,.}
 \eqno(\hbox{C.26})$$
Summing on the first index we find that
$$\eqalign{
\sum_{i=1}^r A_{ij}^{-1} & = {j\over 2}(2r-j)\,, \qquad
j=1,\ldots ,r-1\,, \cr
\sum_{i=1}^r A_{ir}^{-1} & ={r^2\over 2}\,.}
\eqno(\hbox{C.27})$$

\appendix{D}{An explicit bound}

In section 4 we showed abstractly that only finitely many of the
algebras for which there is a single link between the central node and
the affine algebra $C_1$ admit a principal so(1,2) subalgebra. Here we
want to give a more explicit bound for a certain subclass of such
algebras. The subclass of algebras consists of those algebras for
which the central node is linked by precisely one edge to each of the
simple finite Lie algebras in $C_2,\ldots, C_n$, as well as to the
affine node of $C_1$. All algebras are assumed to be simply-laced in
this appendix.

For each $C_p$, let us denote the node that attaches to the central
node by $s_p$. Then $\nu=\sum_{p=2}^{n} \lambda^{(p)}_{s_p}$. Let us
introduce the notation 
$$
X_{j}^{(p)}=(h(\g_1)\, \nu - \rho^{(p)})\cdot\lambda^{(p)}_j=
h(\g_1)\, A^{(p)-1}_{s_p j}-\sum_{i=1}^{r_p}  A^{(p)-1}_{ij}\,.
\eqno(\hbox{D.1})
$$
Next we consider the inequality (A.8) for the case when $j$
corresponds to one of the finite nodes of $C_1$. In terms of 
the inverse Cartan matrix of $\g_1$, this inequality can be written as 
$$\sum_{p=2}^{n} X_{s_p}^{(p)}\le 2h(\g_{1})+1 -{1\over n_j}
\sum_{i=1}^{r_1} A^{(1)-1}_{ij}\,,
\eqno(\hbox{D.2})$$
where $j\in\{1,\ldots,r_1\}$ is arbitrary. Similarly, the
inequality (A.8) for the case when $j$ corresponds to one of the nodes
of $C_2,\ldots,C_n$ is 
$$X_{j}^{(p)}\ge 0 \qquad \hbox{for every $j=1,\ldots, r_p\,, \,$
$p=2,\ldots,n$.}
\eqno(\hbox{D.3})$$
Combining these two equations we therefore find that 
$$
0 \leq  2h(\g_{1})+1 -{1\over n_j} 
\sum_{i=1}^{r_1} A^{(1)-1}_{ij}\,,
\eqno(\hbox{D.4})$$
which only depends on $\g_1$. This inequality implies that 
$r_1\le 15$ for $\g_1=a_{r_1}$ and $r_1\le 16$ for $\g_1=d_{r_1}$. 

Next we observe that for $\g_p=a_{r_p}$, $X_{s_p}^{(p)}\geq 0$ implies
that $r_p+1\le 2 h(\g_1)$. However, the bound of equation (D.3) is
valid for all $j$ and we can find stronger bounds by considering other
values of $j$. In particular, if $s_p\le {r_p\over 2}$, we can choose
$j=2 s_p$, while if $s_p\ge {r_p\over 2}+1$, we can choose
$r_p+1-j=2(r_p+1-s_p)$ and in both cases one finds that equation
(D.3) implies the stronger result $r_p+1\le h(\g_1)$. This in turn 
implies that 
$X_{s_p}^{(p)}\ge {1\over 2} s_p(r_p+1-s_p)\ge {r_p\over 2}$.
This bound applies to all possible values of $s_p$
except the case of $s_p={(r_p+1)\over 2}$ for $r_p$ odd
for which one gets the slightly weaker bound $r_p-1\le h(\g_1)$.
Carrying out a similar analysis for $\g_p=d_{r_p}$ one finds, with the
exception of one value of $s_p$, that one can prove the same bound on
$X_{s_p}^{(p)}$. Hence, apart from these exceptional values of $s_p$,
the bound of equation (D.2) can be written as
$$2+ \sum_{p=1}^n {r_p}\le 
\cases {3-{(r_1+1) (r_1-19)\over 4}\,,\quad {\rm for}\
\g_0=a_{r_1}\,, \cr
5-{(r_1-1) (r_1-18)\over 2}\,,\quad {\rm for}\
\g_0=d_{r_1}\,.}
\eqno(\hbox{D.5})$$
In deriving this equation we have made a suitable subtraction from
both sides so that the the left hand side is just the total rank of
the Kac-Moody algebra $C$. It then follows that the rank of the
Kac-Moody algebra is bounded by $28$ if $\g_0=a_{r_0}$, and by $41$ if 
$\g_0=d_{r_0}$. Actually, with a little bit of extra work one can show
that these bounds also apply to the exceptional values of $s_p$
excluded above.

\footatend\vfill\supereject\immediate\closeout\rfile\writestoppt
\baselineskip=14pt\centerline{{\bf References}}\bigskip{\frenchspacing%
\parindent=20pt\escapechar=` \input refs.tmp\vfill\eject}\nonfrenchspacing

\bye